# Quantum Recurrent Unit: An Effective and Parameter-Efficient Quantum Neural Network Architecture for NISQ Devices

Tzong-Daw Wu [1] and Hsi-Sheng Goan [1,2,3,*]

[1]*Department of Physics and Center for Theoretical Physics, National Taiwan University, Taipei 106319, Taiwan*

[2]*Center for Quantum Science and Engineering, National Taiwan University, Taipei 106319, Taiwan*

[3]*Physics Division, National Center for Theoretical Sciences, Taipei 106319, Taiwan*

* Email: goan@phys.ntu.edu.tw

## Abstract

The rapid growth of modern machine learning models presents fundamental challenges in parameter efficiency and computational resource requirements. We introduce the Quantum Recurrent Unit (QRU), a novel quantum neural network architecture specifically designed to address these challenges while maintaining compatibility with Noisy Intermediate-Scale Quantum (NISQ) devices. QRU leverages quantum controlled-SWAP (C-SWAP) gates to implement an information selection mechanism inspired by classical Gated Recurrent Units (GRUs), enabling selective processing of temporal information through quantum operations. The architecture features a measurement results feedforward scheme with shared parameters across time steps, achieving a constant parameter count and constant per-step effective quantum circuit depth (the depth of the effectively executed quantum circuit incurred over a single QRU time step) regardless of input sequence length. We validate QRU's performance through three progressive experiments under statevector simulation: oscillatory behavior prediction, high-dimensional feature classification of the Wisconsin Diagnostic Breast Cancer (WDBC) dataset, and MNIST handwritten digit classification. Across all tasks, QRU achieves 46.2%–99.7% parameter reduction compared to classical counterparts while achieving comparable performance. Preliminary hardware inference experiments on IBM quantum processors provide initial evidence of execution feasibility on real NISQ devices. The architecture's quantum-native design points toward a candidate quantum-native alternative to classical neural network components, offering a promising direction for parameter-efficient quantum machine learning as hardware capabilities and error mitigation techniques continue to improve.

Keywords: Quantum Machine Learning, Parameter Efficiency, NISQ, Quantum Neural Networks, Recurrent Neural Networks

## 1. Introduction

Modern machine learning development faces a fundamental challenge: As task complexity increases, model sizes continue to expand. Neural networks, serving as the foundational building blocks for Transformers, Convolutional Neural Networks (CNNs), and various types of Recurrent Neural Networks (RNNs), exhibit significant parameter growth as input dimensions increase [1]. This growth not only imposes substantial computational and memory complexity but also limits the model's application in resource-constrained scenarios. In particular, within current complex machine learning architectures, the parameter efficiency of neural network components has become a critical factor constraining further development.

Quantum computing, as an emerging computational paradigm, offers new possibilities for addressing this fundamental challenge. By leveraging quantum mechanical properties such as superposition, entanglement, and interference, quantum systems demonstrate unique computational advantages: they can not only process large amounts of data simultaneously and achieve acceleration in various complex problems [2–5], but more importantly, they exhibit excellent parameter efficiency [6,7], capable of implementing complex function mappings with fewer adjustable parameters [8,9]. Furthermore, quantum computing may potentially lead to lower energy consumption [10,11], offering new insights into addressing the resource efficiency challenges faced by modern machine learning. These advantages of quantum systems collectively point to a possibility: replacing classical neural networks with parameterized quantum circuits to achieve meaningful improvements in parameter efficiency.

However, in the current Noisy Intermediate-Scale Quantum (NISQ) [12] era, translating quantum advantages into practical machine-learning performance remains a significant challenge. Quantum computers are limited by finite quantum capacity [13] and are susceptible to decoherence and noise. While existing quantum machine learning research [14–17] has demonstrated advantages in specific scenarios, it often remains constrained by the number of qubits and circuit depth when handling high-dimensional data [18]. How to effectively utilize quantum resources under these practical constraints has become a core issue in quantum machine learning.

The Quantum Recurrent Unit (QRU) proposed in this paper provides an innovative solution to this problem. Our QRU's design goal is to serve as a quantum alternative to classical neural networks, addressing the fundamental challenge of parameter efficiency and the hardware limitations of the NISQ era. With a recurrent architecture design, our QRUs maintain a constant parameter count and a constant per-step effective quantum circuit depth (the depth of the effectively executed quantum circuit incurred over a single QRU time step) regardless of sequence length. Unlike conventional VQC-based quantum neural networks, which typically rely on classical dense layers to interface with arbitrary input and output dimensions, QRU's recurrent structure enables sequential quantum processing of arbitrary-length sequences without relying on such classical layers. This points toward a direction where classical neural network components could, in principle, be replaced by fully quantum alternatives as hardware matures. This architecture implements efficient information processing through quantum controlled-SWAP (C-SWAP, or Fredkin) gates [19], leveraging both quantum operations and recurrent structure to optimize resource utilization.

Through three representative experiments conducted under statevector simulation, we systematically validate QRU's performance and parameter efficiency: first verifying its fundamental capability as a recurrent architecture through oscillatory behavior prediction tasks, then demonstrating its versatility through two challenging applications - high-dimensional feature classification using the Wisconsin Diagnostic Breast Cancer (WDBC) [20] dataset and image classification using MNIST [21] handwritten digits. Against GRU and MLPs trained in this study, as well as the CNN benchmark [22], QRU achieves 46.2%–99.7% parameter reduction while achieving comparable performance across diverse tasks. These results suggest QRU's parameter-efficiency advantages under the evaluated settings. Preliminary hardware validation on IBM quantum processors is additionally presented in Section 3.5.

This is particularly striking in light of recent comprehensive benchmarking studies that highlight the challenges faced by existing quantum machine learning approaches. Bowles et al. [22] systematically evaluated 12 popular quantum machine learning models across multiple classification tasks, finding that classical models consistently outperformed quantum approaches, with many quantum models struggling particularly on tasks involving amplitude encoding. In their MNIST handwritten digit recognition benchmarks, they found that classical CNNs achieved 96.42% accuracy using 51,841 parameters, while quantum approaches significantly underperformed: Quanvolutional Neural Networks (NNs) reached only 91.41% accuracy, and Wei Net achieved 70.49%. By contrast, our QRU architecture achieves higher accuracy on the same MNIST "3" vs "5" classification task while using 99.7% fewer parameters than the classical CNN, and also outperforms the quantum approaches evaluated in the same study. These results suggest that QRU's quantum-native design may offer a path toward achieving both parameter efficiency and competitive accuracy simultaneously.

## 2. QRU Model

In this section, we introduce our QRU, a quantum recurrent neural network specifically designed for NISQ architectures. To fully understand QRU's design philosophy, we first review relevant theoretical foundations. Given that QRU's information selection mechanism draws inspiration from the Gated Recurrent Unit (GRU), we review the core concepts of classical GRU, the basic principles of quantum computing, and the universal approximation properties of quantum circuits. We then detail QRU's architectural design, from its overall structure to each key component, including its recurrent processing mechanism.

## 2.1 Review of Related Theory

### 2.1.1 Key Aspects of Quantum Computing

Quantum computing leverages quantum state superposition, entanglement, and interference to process information, offering unique computational advantages for specific tasks. As Wu et al. [23] showed, quantum neural networks possess superior expressivity, capable of implementing complex function mappings with relatively few adjustable parameters. However, in the current NISQ era, quantum computers are limited by finite quantum capacity, necessitating careful consideration of qubit-count and circuit-depth constraints when designing quantum algorithms. These hardware limitations are a primary consideration in our QRU design.

### 2.1.2 Update and Reset Mechanisms in Classical GRU

GRU [24] is an efficient variant of recurrent neural networks, with its core idea centered on controlling information propagation through update gates and reset gates. As illustrated in Figure 1, the GRU architecture processes sequential information through three key components: the reset gate, the update gate, and the generation of the candidate hidden state. The reset gate $r_t$ and update gate $z_t$ are computed from the current input $x_t$ and previous hidden state $h_{t-1}$, and then control the generation of candidate hidden state $\tilde{h}_t$ and final hidden state $h_t$. The reset gate $r_t$ determines how much of the previous hidden state information to use when computing the candidate hidden state, while the update gate $z_t$ controls the balance between retaining old information and incorporating new information.

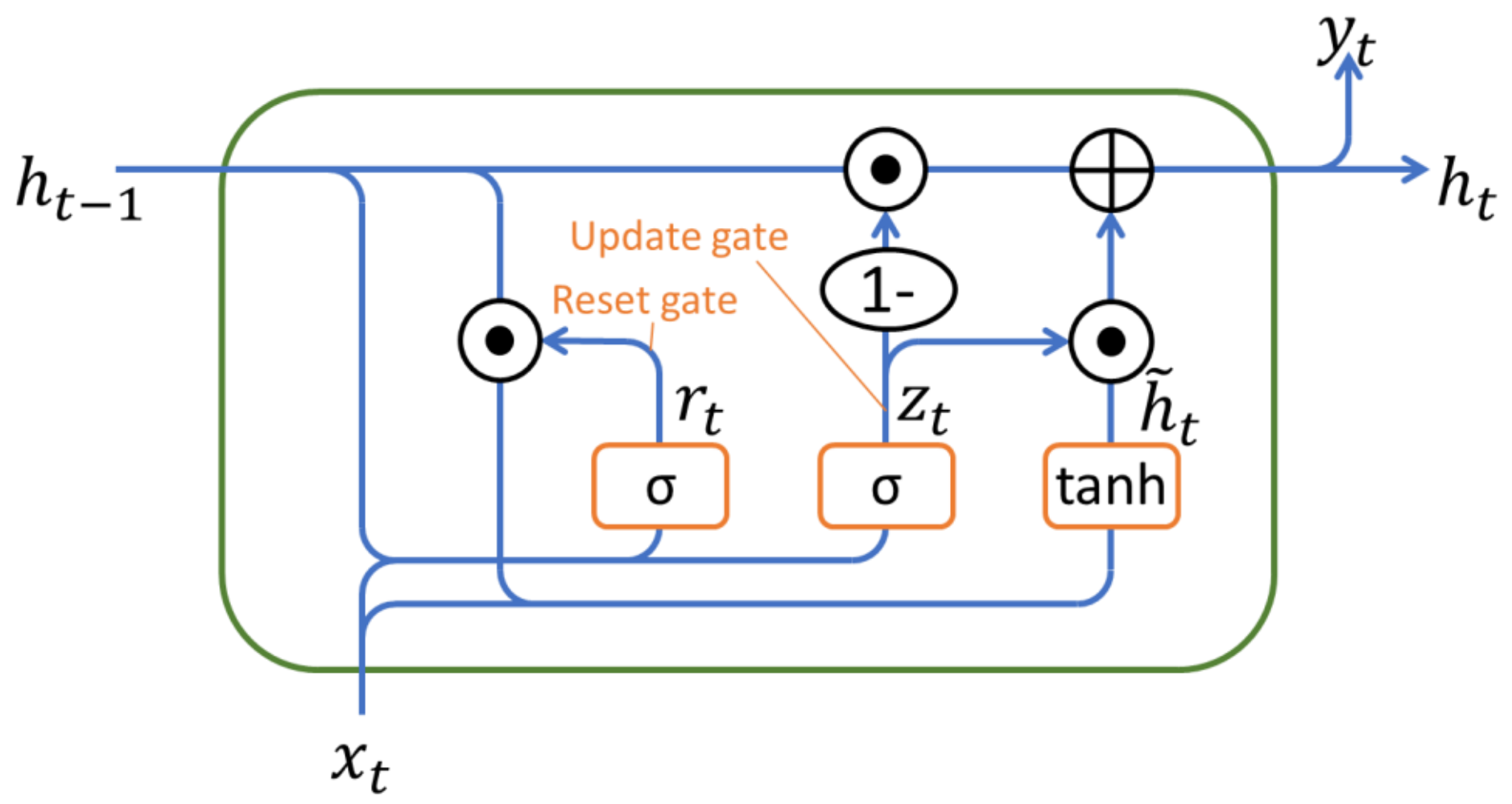

Figure 1: Classical Gated Recurrent Unit (GRU) architecture illustrating the computational flow described in equations (1)–(4). The reset gate controls information flow from the previous hidden state $h_{t-1}$, the update gate determines the mixing ratio between previous and new information, and element-wise operations enable selective information processing, where $\odot$ denotes element-wise multiplication and $\oplus$ denotes addition. This information selection and mixing mechanism inspires QRU's quantum controlled-SWAP operations.

This mechanism can be expressed mathematically as:

$$z_t = \sigma(W_z x_t + U_z h_{t-1} + b_z), \tag{1}$$

$$r_t = \sigma(W_r x_t + U_r h_{t-1} + b_r), \tag{2}$$

$$\tilde{h}_t = \tanh(W_h x_t + U_h (r_t \odot h_{t-1}) + b_h), \tag{3}$$

$$h_t = (1 - z_t) \odot h_{t-1} + z_t \odot \tilde{h}_t, \tag{4}$$

where $\sigma$ is the sigmoid activation function, $\odot$ represents element-wise multiplication, $W$ and $U$ are trainable weight matrices, and $b$ are bias vectors. Notably, equation (4) embodies GRU's core: using the update gate $z_t$ to dynamically adjust the mixing ratio of old and new information. This adaptive information selection mechanism provides important inspiration for subsequent quantum implementation.

### 2.1.3 Universal Approximation Properties of Quantum Circuits

A fundamental theoretical result shows that quantum circuits possess universal approximation capability. Pérez-Salinas et al. [9] demonstrated that even parameterized circuits with a single qubit and data reuploading have universal approximation properties. Schuld et al. [7] further investigated how data encoding strategies affect the expressivity of quantum machine learning models, particularly how appropriate encoding can enhance model expressivity in multi-qubit systems. These findings provide fundamental theoretical support for the design of quantum neural network architectures, demonstrating the feasibility of implementing neural computations with quantum circuits.

### 2.1.4 Current Status and Challenges of Quantum Recurrent Models

Recent research in quantum recurrent neural networks has primarily focused on migrating classical architectures to quantum frameworks. The quantum Long Short-Term Memory (LSTM) proposed by Chen et al. [25] and the quantum GRU architecture developed by Ceschini et al. [26] represent this approach: both replace classical neural network layers with variational quantum circuits (VQC) while retaining the complete classical gating logic — including sigmoid and tanh activation functions and element-wise gating arithmetic — for temporal processing. In these models, VQCs serve as feature extractors that replace classical linear layers, while all temporal decision-making remains classical operations applied to measurement outputs. This hybrid paradigm remains an active area of development, with recent work further developing hybrid QGRU architectures [27], extending such models to predictive maintenance [28] and dynamical systems forecasting [29], and validating related QGRU-based hybrid models through limited real-hardware deployment [30].

This design has a deeper architectural consequence: the classical gating structure of QLSTM and QGRU prevents these models from extending to a fully quantum recurrent form in which hidden state information propagates without intermediate measurement. By contrast, QRU implements the information selection mechanism itself — deciding what to retain from the hidden state and what to incorporate from the new input — natively via C-SWAP operations in the quantum domain, without classical gating arithmetic, a design that in principle supports fully quantum hidden state propagation as hardware matures toward fault-tolerant operation (see Section 2.2.6). QLSTM and QGRU therefore explore whether VQCs can serve as drop-in replacements for classical linear layers within an otherwise unchanged recurrent structure, while QRU additionally explores whether quantum operations can implement the gating mechanism itself, with an extension to support fully quantum hidden state propagation in the fault-tolerant quantum computing era. A comparison with representative QLSTM/QGRU implementations from the literature is provided in Supplementary Section S7.

Beyond such hybrid architectures, more intrinsically quantum recurrent approaches have also been explored. Bausch [31] proposed a quantum RNN (QRNN) using parametrized quantum neurons with polynomial activation functions and amplitude amplification, achieving 90.8%–98.6% accuracy on MNIST binary digit classification, with approximately 1,200 parameters. While demonstrating the feasibility of QRNNs, the architecture's repeat-until-success circuits and amplitude amplification may limit the scalability for longer sequences. Alternative VQC-based approaches [32, 33] maintain quantum hidden states across time steps, demonstrating compact implementations (e.g., 55 parameters in Takaki's work [33]). However, these approaches implement basic recurrent processing without gating mechanisms analogous to GRU or LSTM, potentially limiting their ability to capture long-term dependencies, though more recent work has since introduced a gating mechanism into this line of research [34]. It implements each LSTM/GRU-style gate as a separate parameterized quantum circuit acting on a subset of qubits, with the resulting per-QLSTM/QGRU-gate density matrices combined via further quantum state evolution to update the cell's hidden state, while the hidden state itself remains a quantum state passed directly between cells. QRU instead consolidates this gating logic into a single C-SWAP operation; while this simplifies the gating mechanism structurally, it does not necessarily imply a reduction in circuit depth, as discussed in Section 4.3. Additionally, maintaining quantum coherence throughout sequences, whether in these approaches or in QRU's own extension to support fully quantum hidden state propagation, requires extremely high-fidelity gates to prevent error accumulation across sequential operations and extended coherence times to preserve quantum information—both of which pose practical limitations for NISQ devices.

These limitations have prompted us to reconsider the design approach for quantum recurrent architectures, particularly how to build an architecture that fully leverages quantum advantages through task-specific mechanisms inspired by classical GRU while maintaining NISQ compatibility.

## 2.2 QRU Architecture Design

### 2.2.1 Architecture Overview

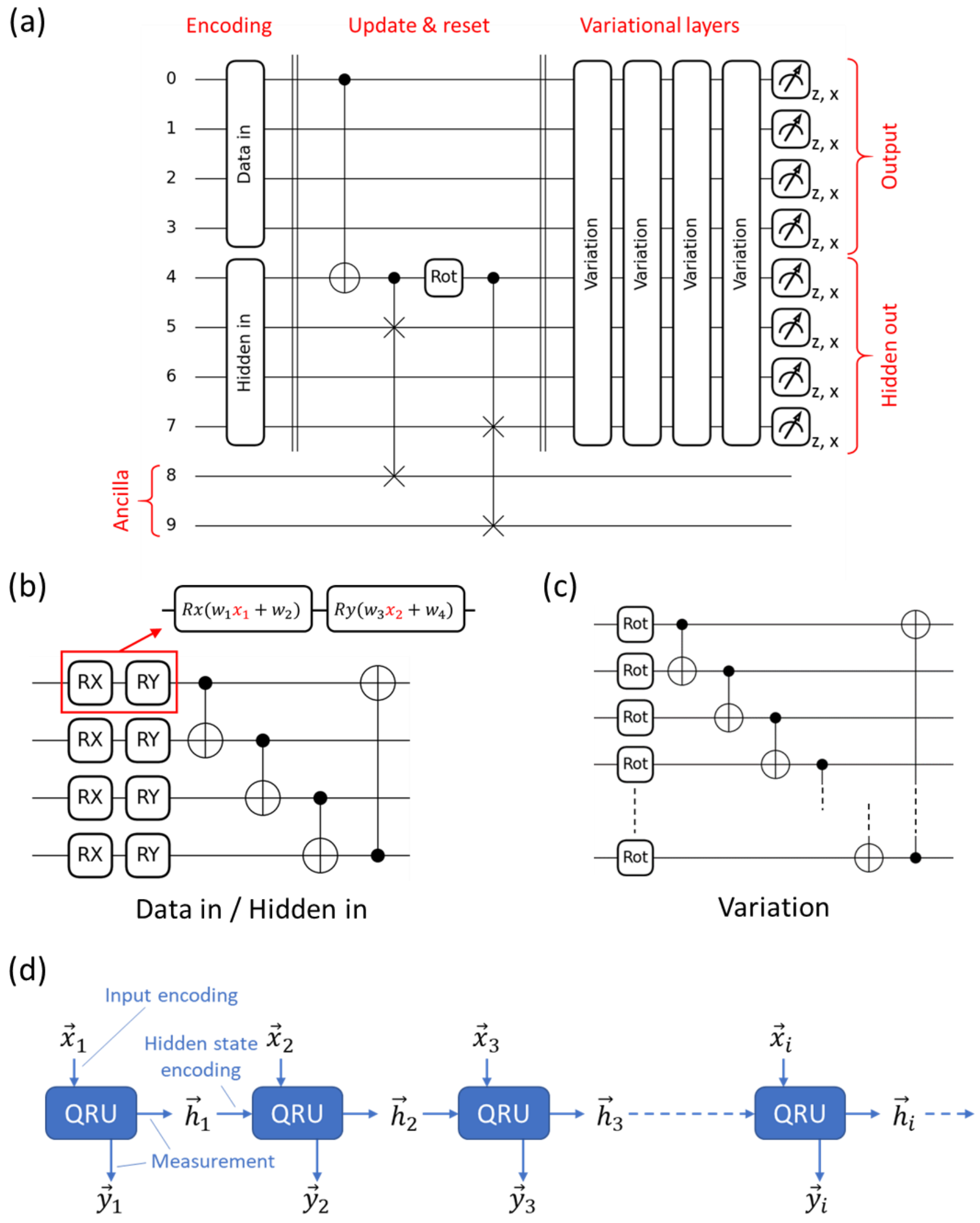

Figure 2: QRU Architecture Diagram. (a) Shows a representative QRU circuit architecture with 4 data qubits, 4 hidden qubits, and 2 ancilla qubits, demonstrating the four main components: input and hidden state encoding, update and reset mechanism, variational layers, and output generation through quantum measurements. Data qubits provide computational outputs, while hidden qubits enable the propagation of temporal information to the next time step. (b) Details the structure of the encoding circuit, showing how rotation gates and entangling CNOT gates are used for input and hidden state encoding. (c) Illustrates the combination pattern of rotation (Rot) gates and entangling CNOT gates in the variational layer. (d) Shows the recurrent processing structure in which multiple QRUs process sequential data, with hidden states propagated between time steps through measurement results feedforward schemes to enable modeling of temporal dependencies.

Figure 2 illustrates QRU's architectural design, which includes four main components and their recurrent processing structure:

- Input and Hidden State Encoding: Converting classical data into rotation angles for quantum states using angle encoding schemes. (Section 2.2.2)

- Update and Reset Mechanism: Using quantum controlled-SWAP (C-SWAP) operations to mimic classical GRU's update and reset logic through selective information processing. (Section 2.2.3)
- Variational Layers: Composed of parameterized quantum gates for approximating complex functions. (Section 2.2.4)
- Output and Hidden State Update: Producing model outputs and updating hidden states through quantum measurements in Z and X bases. (Section 2.2.5)
- Recurrent Processing Structure: Connecting multiple QRUs to process sequential data with a measurement results feedforward scheme [35,36], enabling state propagation and temporal dependency modeling with a constant per-step effective quantum circuit depth regardless of sequence length. (Section 2.2.6)

### 2.2.2 Input and Hidden State Encoding

In the encoding phase, as shown in Figure 2(b), QRU adopts an efficient angle encoding scheme for both input data and hidden states. This encoding converts values ($x_1$, $x_2$) into rotation angles for gates:

$$\begin{aligned} R_x(\theta_1) &= R_x(w_1 x_1 + w_2), \\ R_y(\theta_2) &= R_y(w_3 x_2 + w_4), \end{aligned} \tag{5}$$

where $w_1$, $w_2$, $w_3$, $w_4$ are trainable parameters. Each qubit can encode two dimensions of information through $R_x$ and $R_y$ gates, and quantum entanglement is introduced through circular CNOT entanglement topology, implementing correlations between qubits. This design fully utilizes the state space of qubits while maintaining a moderate depth in the encoding circuit.

### 2.2.3 Update and Reset Mechanism

QRU implements an information selection mechanism similar to GRU using quantum controlled-SWAP (C-SWAP) or Fredkin gates [19]. As discussed in Section 2.1.2 regarding the GRU update mechanism, the QRU also needs to balance retaining information from the previous hidden state and incorporating new information.

As shown in Figure 3, a C-SWAP operation allows the system to selectively swap the states of two target qubits based on the state of the control qubit:

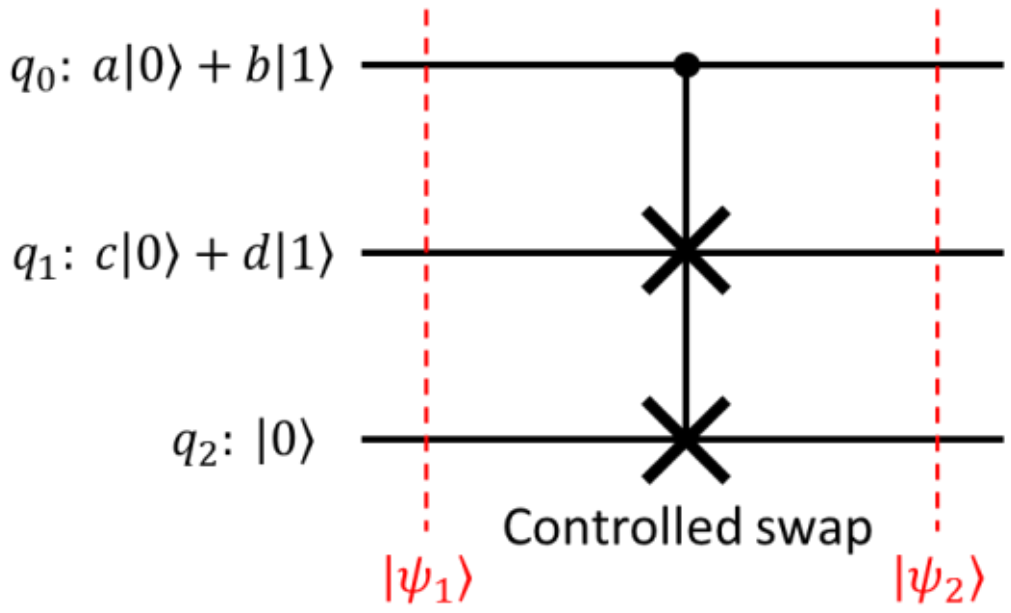

Figure 3: Update and Reset Mechanism Diagram of QRU, showing how C-SWAP operations achieve selective state exchange to accomplish information-selective updates.

This quantum mechanism can be described by state evolution:

$$|\psi_1\rangle = ac|000\rangle + ad|010\rangle + bc|100\rangle + bd|110\rangle$$
$$\downarrow$$
$$|\psi_2\rangle = ac|000\rangle + ad|010\rangle + bc|100\rangle + bd|101\rangle. \quad (6)$$

The corresponding expectation value of the Pauli-Z operator on qubit q1 evolves as:

$$\langle\psi_1|\sigma_z(q_1)|\psi_1\rangle = |c|^2 - |d|^2$$
$$\downarrow$$
$$\langle\psi_2|\sigma_z(q_1)|\psi_2\rangle = |a|^2(|c|^2 - |d|^2) + |b|^2, \quad (7)$$

where $\sigma_z(q_1)$ represents the Pauli-Z operator acting on qubit q1. This evolution process implements GRU's update mechanism through quantum operations in a more integrated way. The control qubit $q_0$, through coefficients $|a|^2$ and $|b|^2$, not only determines the mixing ratio but also participates in introducing new information during the quantum evolution. The target qubit $q_1$ undergoes state changes through the C-SWAP operation, corresponding to the hidden state transformation. The ancilla qubit $q_2$ enables this selective state exchange, making the controlled information update possible.

In QRU's implementation, information selection operates at two levels: the C-SWAP operation itself provides fine-grained mixing control through quantum amplitudes, while the selective application to specific qubits enables coarse-grained architectural control. Not all qubits participate in C-SWAP operations; some undergo controlled swaps, while others retain their information unchanged, mimicking GRU's selective update behavior, in which different hidden state elements experience different degrees of modification.

This quantum-native approach leverages the properties of quantum superposition states to simultaneously achieve information update and fusion, unlike GRU's sequential process of first generating candidate states and then mixing them with previous states. An ablation study isolating the contribution of the C-SWAP mechanism is provided in Supplementary Section S4.2.

### 2.2.4 Variational Layers

The design of variational layers, as shown in Figure 2(c), builds upon the theoretical basis of quantum circuits' approximation properties discussed in Section 2.1.3. Each layer contains multiple rotation gates with independent trainable parameters, serving as the primary information processing component after the QRU's update mechanism. While quantum circuits theoretically possess universal approximation capability, our design focuses on achieving practical processing power under NISQ device constraints through its circuit structure.

### 2.2.5 Output and Hidden State Update

After processing through variational layers, each qubit is measured in the Z and X bases at the end of each QRU. The measurement results for all qubits are split into two parts, serving two crucial purposes: generating output for the current time step and updating hidden states to feed forward to the next QRU. While the output provides the model's prediction or classification results, the hidden states fed forward to the following QRU play a key role in maintaining temporal information processed by QRU's update mechanism, which is implemented through C-SWAP operations, providing an information foundation for the next time step's processing. By performing measurements in different bases, we can obtain richer information from quantum states, thereby ensuring effective transmission of information across the time dimension. An ablation comparison between dual-basis (Z+X) and Z-only measurement is provided in Supplementary Section S4.1.

### 2.2.6 Recurrent Processing Structure

As shown in Figure 2(d), multiple QRUs are connected in sequence through a measurement results feedforward scheme to process sequential data, forming a recurrent architecture. Each QRU processes input data $\vec{x}_i$ and receives hidden states $\vec{h}_i$ from the previous time step, generating both output vectors $\vec{y}_i$ and updated hidden states for subsequent processing. At the end of each QRU, quantum measurements are performed on each of the qubits. The measurement results are categorized into two parts: one as system output and the other as hidden states fed forward to the next QRU.

The use of a measurement results feedforward scheme between quantum circuits has been explored for constructing quantum NN architectures. Lai [35] demonstrated that this approach to quantum depth scaling can introduce a classical activation-like nonlinearity between the measurement and re-encoding, with independent parameters per layer. While the QRU model adopts a similar measurement-and-re-encoding connection pattern, it applies this pattern to temporal processing, using shared parameters across time steps, enabling recurrent computation and parameter efficiency.

Specifically, the hidden state propagation follows a specific encoding process: hidden qubits are measured in both the Z and X bases to extract expectation values, which are then fed forward and encoded into the next QRU using the same angle encoding scheme shown in Figure 2(b). This ensures that temporal information processed by the current QRU's update mechanism is effectively transmitted to subsequent processing steps.

Through this design, the QRU model's architecture, with measurement results fed forward and shared parameters across different QRUs, ensures that both the total parameter count and the effective quantum circuit depth (the depth of the effectively executed quantum circuit) remain constant, determined solely by a single QRU, regardless of sequence length. This enables processing of arbitrary-length sequences while maintaining NISQ compatibility, with the effectively executed quantum circuit depth determined by a single QRU time step.

This constant per-step effective quantum circuit depth reflects an active design choice for NISQ compatibility rather than an architectural constraint. The QRU architecture inherently supports a fully quantum recurrent extension: by retaining the hidden qubits without measurement and resetting the ancilla qubits at the start of each time step, hidden state information can in principle propagate as a quantum state across time steps without classical collapse. This extension preserves the same C-SWAP-based gating logic and requires only limited changes to the circuit structure. The effective depth of executing the quantum circuit for fully quantum hidden state propagation, however, increases linearly with the sequence length, in direct contrast to the constant per-step effective quantum circuit depth regardless of sequence length in the QRU model's NISQ-compatible architecture with the measurement results feedforward scheme. The fully quantum recurrent extension will become practical as hardware matures toward fault-tolerant operation that can reliably execute long quantum circuits.

We distinguish two senses in which a model can be quantum-recurrent: weak quantum recurrence, in which information propagates step by step through a shared update function — for example, $h_i = f(h_{i-1}, x_i; \theta)$ with parameters $\theta$ shared across steps, following the notation used elsewhere in this section — regardless of whether $h_i$ is represented classically or as a quantum state; and strong quantum recurrence, which additionally requires that the hidden state persists as a coherent quantum state across steps, without an intervening measurement. QRU's current implementation presented in this paper satisfies weak quantum recurrence: the measurement results feedforward scheme is, under this definition, a specific realization of the update function $f$, not a departure from the recurrent architecture class. The fully quantum extension described above would additionally satisfy strong quantum recurrence once hardware permits sustained coherence across time steps.

Beyond scalability in the depth of quantum circuits in time-step sequences, this architecture addresses another key challenge in quantum machine learning: enabling the QRU model to handle arbitrary input and output dimensions through sequential processing. This dimensional flexibility makes the model adaptable to diverse task requirements while operating within NISQ device constraints.

One practical constraint of the current implementation is the inter-step latency inherent in the measurement results feedforward scheme: measurement outcomes from step $t$ must be processed and re-encoded before the circuit for step $t$+1 can execute, so sequential time steps cannot be pipelined in parallel on a single quantum processor even with a larger qubit count, and total wall-clock time scales linearly with sequence length. This constraint is shared by all mid-circuit measurement feedforward architectures on current hardware, and is discussed further in the context of hardware validation in Section 3.5.

# 3. Experimental Validation

## 3.1 Experimental Design Overview

This study validates QRU's performance and universality through three progressively more complex experiments. The first experiment focuses on time series prediction, selecting oscillatory behavior prediction as basic validation, which is not only a typical application of RNNs but also an ideal test scenario for evaluating QRU's temporal modeling capabilities. The second experiment extends to high-dimensional feature processing, using the WDBC dataset classification task to validate QRU's effectiveness in handling high-dimensional data. The third experiment further extends to the image recognition domain, using the MNIST handwritten digit recognition task to demonstrate QRU's performance in handling more complex visual data.

All simulation experiments use the PennyLane [37] quantum machine learning framework combined with the JAX [38] backend. Although these experiments simulate quantum circuits on classical computers, we take fundamental quantum computing properties and resources into consideration in the design, such as qubit count usage and circuit depth control. The choice of the simulation environment is primarily due to the extensive iterative training requirements of machine learning, where each model training run involves thousands of iterations, making the implementation on current real quantum computers, with limited availability and access times, expensive and impractical for such intensive computational demands. All performance results reported in Sections 3.2–3.4 are therefore obtained under statevector simulation. The experimental methodology and training configurations adopted across all three experiments are described below. We also present preliminary hardware validation experiments separately in Section 3.5.

While these three experiments differ in application scenarios, they adopt similar training strategy frameworks. For optimization, all experiments use the Adam optimizer and implement dynamic learning rate adjustment mechanisms. Specifically, the system records training loss values at fixed intervals (10 or 100 epochs, depending on experimental settings). When observing that three consecutive recorded loss values are all higher than the fourth-last recording, the system automatically halves the learning rate and reverts to the model parameter state with the lowest loss value during that training period to continue training. This adaptive mechanism effectively prevents training instability issues caused by excessive learning rates, ensuring robust model convergence to optimal solutions.

To prevent overfitting, each experiment employs an early stopping mechanism, with stopping conditions adjusted to different task characteristics and data structures. Similarly, the choice of loss functions varies based on task objectives.

Regarding model architecture, the QRU structures in all three experiments are based on the basic architecture described in Section 2 but differ in specific implementation details to optimize the performance-efficiency balance for each task. The hidden state configurations range from 4 to 5 qubits, each measured in both the Z and X bases to extract expectation values (yielding 8- to 10-dimensional hidden state representations), and the variational layer designs range from single-parameter to three-parameter rotation gates across 3 to 4 measurement-result feedforward circuit layers. These configurations were determined through preliminary experiments to achieve stable performance while maintaining parameter efficiency for the specific task complexity. The detailed specifications for each experiment are presented in the respective experimental sections.

With these flexible configurations, we adopt a progressive validation strategy. Although QRU's ultimate goal is to serve as a quantum alternative to classical neural networks, as a first step, we need to validate its basic capability as a recurrent architecture. Therefore, in time series prediction tasks, we compare against GRU and LSTM, as these mature classical RNN models provide good benchmarks. After confirming QRU's recurrent processing capabilities, we further validate its performance on more challenging tasks: high-dimensional feature classification (WDBC) and visual data processing (MNIST), demonstrating QRU's potential as a universal neural network alternative. This progressive experimental design allows us to systematically evaluate QRU's spectrum from basic capabilities to universal applications.

## 3.2 Oscillatory Behavior Prediction: Basic Temporal Capability Validation

### 3.2.1 Experimental Method

The first experiment aims to validate QRU's basic capabilities for modeling and predicting time-series data. By simultaneously handling Simple Harmonic Oscillation and Damped Oscillation with a single model, we evaluate the model's ability to maintain temporal memory within a learned dynamical regime across different oscillation patterns. In particular, the introduced self-feedback prediction mechanism further demonstrates the model's potential in handling long-term temporal dependencies.

The experiment uses two sets of time-series data, each containing 150 data points, comprising sequences of simple harmonic oscillation and damped oscillation. The first 100 data points from each dataset are used for joint training of the model, while the remaining 50 points serve as the test set to evaluate the model's predictive performance. During training, a step-by-step prediction approach is adopted, in which the model receives a single data point as input and predicts the value for the next time step, learning to handle both oscillation patterns simultaneously.

Given that this task involves one-dimensional time-series data, we implement a QRU with 5 hidden qubits and 4-layer variational circuits— a configuration determined through preliminary experiments following the approach outlined in Section 3.1. Specifically, the QRU uses a single qubit with a single Rx rotation gate for data encoding; the hidden state is implemented using 5 qubits to realize a 10-dimensional state representation; the variational layers are constructed from 4 layers of quantum circuits, each utilizing U2($\varphi$, $\lambda$) rotation gates—two-parameter single-qubit rotations implementing Rz($\varphi$)Ry($\pi/2$)Rz($\lambda$)—with $\varphi$ and $\lambda$ as trainable parameters. These layers then generate prediction output by measuring the Z-basis expectation value of the first qubit. The overall architecture contains 72 trainable parameters. The complete circuit architecture and a per-component parameter breakdown are given in Supplementary Figure S1 and Section S1.4, respectively.

Model training uses Mean Squared Error (MSE) as the loss function. To evaluate the model's capacity to jointly learn and distinguish multiple temporal patterns, a joint training strategy is adopted, enabling the same model to learn the dynamic characteristics of both simple harmonic oscillation and damped oscillation simultaneously. To prevent overfitting, an early stopping mechanism is implemented, terminating training when the decrease in training loss recorded every 100 epochs is less than 1% for three consecutive recordings.

For comprehensive statistical analysis, we conducted 50 independent experiments for different parameter configurations of each model type. Specifically, QRU uses a fixed configuration of 72 parameters, classical GRU tests multiple configurations ranging from 11 to 441 parameters, and LSTM experiments with different scales ranging from 101 to 491 parameters. Each configuration independently executes 50 training runs, each using the same model structure and training strategy but with different parameter initializations. This experimental design, with a systematic parameter sweep, allows us to comprehensively analyze the impact of parameter scale on model performance. To ensure robustness of the reported statistics, outliers in the test results are identified using Tukey's fences criterion (values below Q1 − 1.5 × IQR or above Q3 + 1.5 × IQR, where Q$i$ indicates the $i$th quartile the data points belong to, and IQR=Q3-Q1 stands for Interquartile Range) and reported separately to characterize the influence of runs that may have converged to poor local minima. The primary comparisons are based on the complete results including all runs.

### 3.2.2 Results Analysis and Comparison

As shown in the detailed statistical metrics in Table 1, QRU demonstrates consistent performance in handling both oscillation patterns simultaneously with a single set of parameters across 50 independent experiments. During the training phase, the model achieved a mean MSE of 2.14E-04 with a standard deviation of 2.06E-04, indicating stable learning of both patterns. The training loss curves from 50 independent runs are shown in Supplementary Figure S2, indicating a consistent downward trend throughout training. The mean MSE in the testing phase was 1.94E-02 with a standard deviation of 4.22E-02; after removing statistical outliers, the mean and standard deviation decreased to 2.49E-03 and 3.06E-03, respectively, reflecting the influence of a small number of high-error runs among the 50 independent experiments.

Table 1: QRU Model Performance Statistics in Training and Testing Phases (Including Original Data and Results After Outlier Removal)

| | Train | Train (w/o outlier) | Test | Test (w/o outlier) |
|---|---|---|---|---|
| **Mean** | 2.14E-04 | 1.65E-04 | 1.94E-02 | 2.49E-03 |
| **Median** | 1.64E-04 | 1.55E-04 | 1.78E-03 | 9.67E-04 |
| **Std. dev.** | 2.06E-04 | 7.84E-05 | 4.22E-02 | 3.06E-03 |

The prediction capability is illustrated in Figure 4, which presents a representative experimental result. In the simple harmonic oscillation prediction task (Figure 4(a)), the QRU accurately captured the periodicity and amplitude characteristics of the oscillation, with the prediction curve (green line) closely matching the test data (red triangles), indicating that the model successfully learned the fundamental properties of harmonic motion. In the damped oscillation prediction task (Figure 4(b)), the model performed equally well, accurately predicting not only the oscillation frequency but also successfully simulating the amplitude decay trend. Notably, even in later prediction periods, the model's output remains consistent with the actual data.

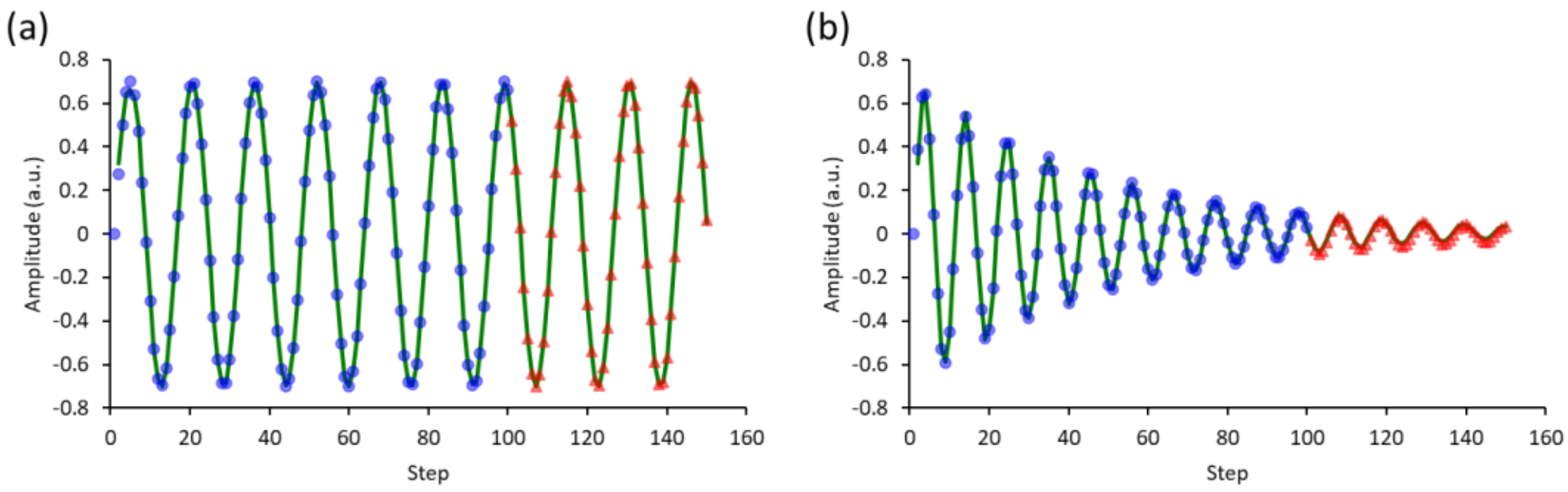

Figure 4: Comparison of model predictions with actual data. (a) Simple harmonic oscillation and (b) damped oscillation prediction results, where blue dots represent training data, red triangles represent test data, and the green line represents model predictions.

To evaluate the parameter efficiency of QRU, we first compared it with classical models using similar parameter counts. As shown in Table 2, with 72 parameters, QRU (test MSE: 1.94E-02) shows substantial improvement in test MSE over both GRU (77 parameters, MSE: 1.60E-01) and LSTM (101 parameters, MSE: 7.19E-01). Moreover, QRU's test standard deviation (4.22E-02) is also smaller than GRU (2.13E-01) and LSTM (1.41E+00), indicating that its predictions are not only more accurate but also more stable.

Table 2: Training and Testing MSE Statistics for Three Models with Similar Parameter Counts

| | QRU (72) | GRU (77) | LSTM (101) |
|---|---|---|---|
| **Train Mean** | 2.14E-04 | 2.58E-03 | 6.08E-05 |
| **Train (w/o outlier)** | 1.65E-04 | 4.31E-05 | 6.08E-05 |
| **Test Mean** | 1.94E-02 | 1.60E-01 | 7.19E-01 |
| **Test (w/o outlier)** | 2.49E-03 | 1.05E-01 | 2.14E-01 |
| **Test Std. dev.** | 4.22E-02 | 2.13E-01 | 1.41E+00 |

Given QRU's competitive performance with minimal parameters, we conducted parameter-expansion experiments on classical models to determine how many parameters they would need to achieve comparable performance levels. As shown in Table 3, GRU requires expansion to 307 parameters to achieve a comparable full test mean MSE to the 72-parameter QRU (GRU-307: 1.80E-02 vs QRU: 1.94E-02). After removing statistical outliers, which accounts for runs that may have converged to poor local minima due to stochastic initialization, GRU requires 197 parameters to match QRU's performance (GRU-197: 2.22E-03 vs QRU: 2.49E-03), while LSTM, even expanded to 491 parameters, still cannot surpass QRU's performance. This parameter requirement is further illustrated in Figures 5 and 6. As shown in Figure 5, during the training phase, classical models exhibit performance improvements with increasing parameter counts, though the improvement margin gradually diminishes. The testing phase results in Figure 6 further confirm that GRU and LSTM require significantly more parameters to approach QRU's performance level.

Table 3: Performance Statistics Comparison of Models After Parameter Expansion

| | **QRU (72)** | **GRU (197)** | **GRU (307)** | **LSTM (491)** |
|---|---|---|---|---|
| **Train Mean** | 2.14E-04 | 3.43E-04 | 6.07E-03 | 3.42E-05 |
| **Train (w/o outlier)** | 1.65E-04 | 3.37E-05 | 3.36E-05 | 3.34E-05 |
| **Test Mean** | 1.94E-02 | 7.28E-02 | 1.80E-02 | 4.69E-01 |
| **Test (w/o outlier)** | 2.49E-03 | 2.22E-03 | 1.62E-03 | 1.64E-01 |
| **Test Std. dev.** | 4.22E-02 | 3.85E-01 | 4.64E-02 | 1.04E+00 |

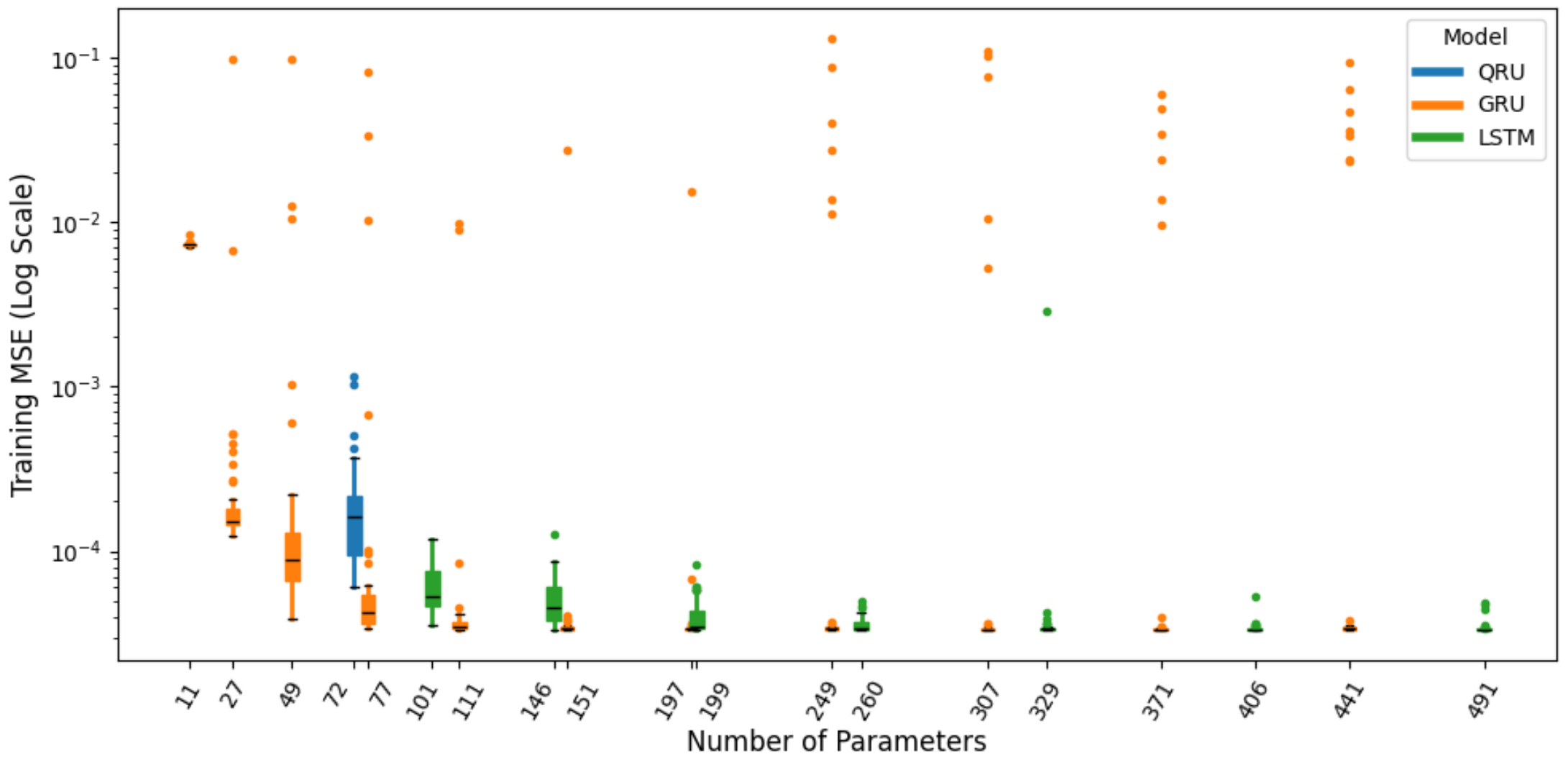

Figure 5: Training performance comparison of three models under different parameter scales. Box plots show the MSE distributions for QRU (blue), GRU (orange), and LSTM (green), with 50 independent training runs for each parameter

configuration. The X-axis shows the number of model parameters, and the Y-axis shows the corresponding MSE values (in log scale). Each box plot represents the statistical distribution from these training runs.

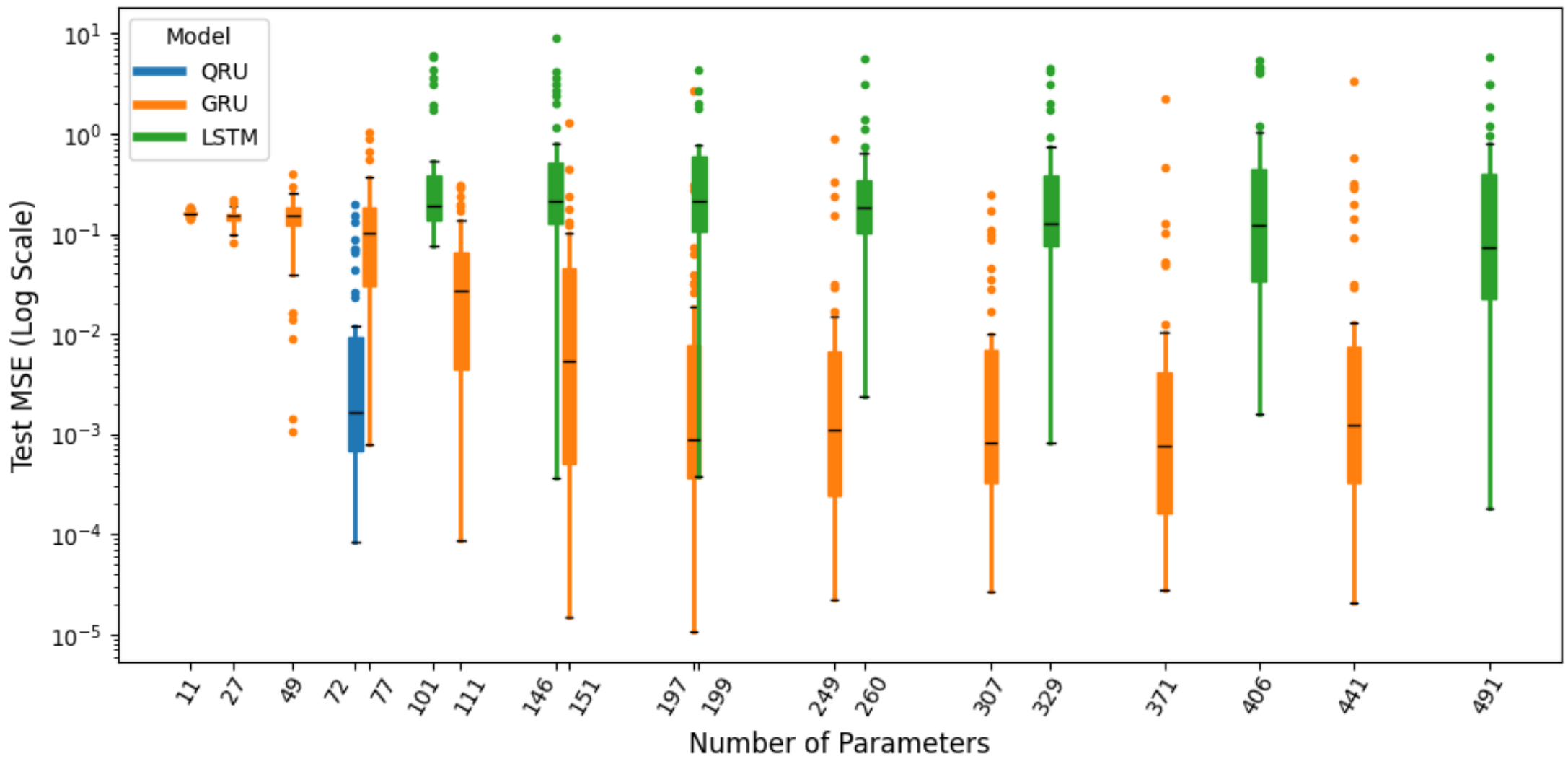

Figure 6: Model performance distribution on the test set. Using the same visualization approach as in Figure 5, the MSE distributions of three models are shown during the testing phase. Each box plot represents the distribution of test results from 50 independent training runs for a specific parameter configuration, clearly showing the impact of different configurations on model generalization.

## 3.3 WDBC Classification: High-Dimensional Feature Processing Capability Validation

### 3.3.1 Experimental Method

This experiment aims to validate QRU's capability in handling high-dimensional feature classification problems. We conduct experiments using the WDBC dataset from the UCI Machine Learning Repository [39]. The dataset contains 30 features extracted from digital images of breast tumor cell nuclei, used to distinguish between benign and malignant tumors.

The dataset comprises 569 samples, including 357 benign tumors and 212 malignant tumors. Given the relatively small dataset, we adopt the Leave-One-Out Cross-Validation (LOOCV) strategy for experimental evaluation to make full use of the limited data. In the data preprocessing phase, we apply Min-Max normalization to all features, scaling their values to the [0,1] interval. During actual processing, the model sequentially processes these 30 standardized feature values.

The model architecture adopts a sequential processing strategy similar to the oscillation prediction experiment, treating the 30-dimensional features as a one-dimensional sequence of length 30. For this classification task, we implement a QRU with 4 hidden qubits and 3-layer alternating rotation gates, consistent with the configuration approach discussed in Section 3.1. Specifically, the input encoding uses a single qubit with an Rx rotation gate for feature encoding; the hidden state uses 4 qubits to implement an 8-dimensional state representation; the variational layers are constructed with three layers of rotation gates, adopting an alternating design pattern: the first layer uses Rx gates, the second layer uses Ry gates, and the third layer uses Rx gates again, with each rotation gate containing 1 trainable parameter. The model's final output takes the Z-basis expectation value of the first qubit, and we map the output range from [-1,+1] to [-*s*,+*s*] through a trainable parameter *s* to optimize binary classification effects. The overall model contains 35 trainable parameters. Detailed implementation and a per-component parameter breakdown are provided in Supplementary Figure S3 and Section S2.4, respectively.

The training process employs batch training with a batch size of 20. Since this task focuses on binary classification of tumors into benign and malignant, we use binary cross-entropy as the loss function:

$$loss = -\frac{1}{N}\sum_{i=1}^{N}(y_i \cdot \log(\hat{y}_i) + (1 - y_i) \cdot \log(1 - \hat{y}_i)) \tag{8}$$

where $y_i$ is the true label for the *i*-th sample (0 for malignant, 1 for benign), $\hat{y}_i$ is the model's predicted probability of the *i*-th sample being benign, and *N* is the number of samples.

To prevent overfitting, we implemented an early stopping mechanism. The system records training loss every 10 epochs, and when the decrease in loss values for three consecutive recordings is less than 1%, the training process is terminated. This design helps stop training when model performance stabilizes, avoiding overfitting to training data.

### 3.3.2 Results Analysis and Comparison

As shown in Table 4, the QRU model shows strong performance on the WDBC classification task. In 569 LOOCV experiments, the model achieved a test accuracy of 96.13%, with a stable training process (training accuracy 97.97%, standard deviation 0.95%). The confusion matrix shows that 347 out of 357 benign samples were correctly identified, and 200 out of 212 malignant samples were correctly identified, with only 10 false positives and 12 false negatives, corresponding to a sensitivity of 94.34% and a specificity of 97.20%, with malignant samples treated as the positive class. These results demonstrate QRU's effectiveness on the WDBC classification task, with performance metrics detailed in Table 4. Training loss curves across all LOOCV folds are provided in Supplementary Figure S4, confirming stable optimization throughout training.

Table 4: Detailed Performance of QRU Model on WDBC Classification Task

| Metric Type | Metric | Value |
|---|---|---|
| **Classification Performance** | Test Accuracy | 96.13% (547/569) |
| | Training Accuracy | 97.97% (±0.95%) |
| **Confusion Matrix** | True Positive (TP) | 200 |
| | True Negative (TN) | 347 |
| | False Positive (FP) | 10 |
| | False Negative (FN) | 12 |
| **Derived Metrics** | Sensitivity | 94.34% |
| | Specificity | 97.20% |
| | Precision | 95.24% |
| | F1-Score | 94.79% |
| **Training Process** | Mean Epochs | 658.23 (±260.81) |
| | Epoch Range | 80–2040 |

QRU achieves this performance level using 35 parameters. As shown in Table 5, even compared to neural network benchmark results reported in the UCI Machine Learning Repository (accuracy 92.31%, range 87.41%–96.50%), QRU demonstrates consistent performance. Particularly noteworthy: while some studies achieved higher classification performance through feature engineering, we specifically selected neural network baseline results using the original features (without feature selection or extraction) from these studies for comparison. Compared to classical neural networks using more parameters, such as Guo and Nandi's Multilayer Perceptron (MLP) [40] (65–449 parameters†, 96.21%), Aalaei et al.'s Artificial Neural Network (ANN) [41] (167 parameters, 96.50%), and Rani et al.'s Deep Neural Network (DNN) [42] (1,601 parameters†, 95.32%), QRU achieves comparable performance levels (96.13%) with fewer parameters (35).

Table 5: Different Models' Performance on WDBC Test Set

| Model | Parameters | Test Accuracy | Additional Metrics |
|---|---|---|---|
| QRU Model (This study) | 35 | 96.13% | Sen: 94.34%, Spec: 97.20%, Prec: 95.24%, F1: 94.79% |
| ***Classical counterparts trained in this study*** | | | |
| MLP (2 hidden neurons) | 65 | 96.84% | Sen: 95.28%, Spec: 97.76%, Prec: 96.19%, F1: 95.73% |
| MLP (14 hidden neurons) | 449 | 97.36% | Sen: 96.23%, Spec: 98.04%, Prec: 96.68%, F1: 96.45% |
| MLP (WEKA default) | 530 | 97.01% | Sen: 96.23%, Spec: 97.48%, Prec: 95.77%, F1: 96.00% |
| ***Reported in the literature*** | | | |
| MLP [Guo & Nandi][40] | 65–449† | 96.21% (±1.73%) | - |
| ANN [Aalaei et al.][41] | 167 | 96.50% | Sen: 98.20%, Spec: 96.00% |
| MLP [Salama et al.][43] | - | 96.66% | - |
| DNN [Rani et al.][42] | 1,601† | 95.32% | Sen: 95.08%, Spec: 95.45%, Prec: 92.06%, F1: 93.54% |
| NN [UCI][39] | - | 92.31% (87.41%–96.50%) | Prec: 91.64% (86.51%–95.97%) |

†Estimated, as the original papers do not fully specify the network architecture; all other parameter counts are exact. Models trained in this study used the same protocol as QRU (Section 3.3.1), in a single LOOCV run. The MLPs with 2 and 14 hidden neurons follow the architecture of [40] (one sigmoid hidden layer and a linear output layer, whose size is not reported; a single linear output neuron is used here) and span the hidden-layer sizes considered there (2–14). The third MLP follows the default architecture of the WEKA software [44] used in [43], as [43] does not report its settings: (30 attributes + 2 classes)/2 = 16 sigmoid hidden neurons and one sigmoid output neuron per class.

Because some of these published parameter counts are estimated and the training protocols differ from ours, we also trained MLPs with the same preprocessing, LOOCV protocol, and training settings as QRU (Section 3.3.1; Table 5). With 2 and 14 hidden neurons, the smallest and largest hidden layers considered in [40], they reach test accuracies of 96.84% with 65 parameters and 97.36% with 449 parameters; in the default configuration of the WEKA software [44] used in [43], an MLP reaches 97.01% with 530 parameters. QRU's test accuracy is 0.71–1.23 percentage points lower than these MLPs, with 46.2%–93.4% fewer parameters (Table 11).

To further examine whether QRU's classification performance depends on a particular feature ordering, we evaluated the model across five independently randomized feature orderings, each paired with a different stratified dataset split. The standard deviation across the five per-ordering mean accuracies is 0.97% (detailed results in Supplementary Section S4.3), indicating that QRU's performance is robust to the choice of input feature ordering and does not rely on any assumed sequential structure in the WDBC feature set.

## 3.4 MNIST Digit Classification: Image Data Processing Capability Validation

### 3.4.1 Experimental Method

This experiment aims to validate QRU's capability in handling high-dimensional image data. Following the benchmark tests in Bowles et al. [22], we specifically focus on the digit recognition task for "3" and "5" using their MNIST dataset. This pair was chosen in [22] because it is among the hardest digits to distinguish, providing a challenging test scenario for model evaluation. Through converting two-dimensional image data into sequential input, this experiment also demonstrates QRU's flexibility in handling non-sequential data.

The experiment uses the 8×8-pixel version of the MNIST dataset provided by [22], in which the original 28×28 MNIST images were resized using bilinear interpolation to preserve spatial correlation structure. The CNN benchmark in [22] also uses this same 8×8 version, ensuring that all models are compared under identical input conditions. Hyperparameters of the CNN, including convolutional kernel size, were selected via independent grid search for each resolution tested in [22] (4×4, 8×8, 16×16, and 32×32), rather than fixed for a specific resolution; for the 8×8 configuration used here, this ensures the CNN baseline reflects its best-performing configuration at this resolution.

The dataset contains 7,141 samples of "3" and 6,313 samples of "5". We employ 7-fold stratified cross-validation for experimental evaluation, ensuring that each fold preserves the original dataset's class proportions. In each fold's training process, we further randomly extract 1/6 of the training data as a validation set, ultimately forming a 5:1:1 split ratio for training, validation, and test sets.

The implemented QRU structure is specifically designed for high-dimensional data processing. For this task, we implement a QRU with 4 hidden qubits and 4-layer Rot-based variational circuits. Specifically, the input encoding uses 4 qubits, capable of processing 8 features per time step; the hidden state similarly uses 4 qubits, providing 8-dimensional state representation capability; the variational layer uses Rot gates—three-parameter single-qubit rotations implementing Rz($\omega$)Ry($\vartheta$)Rz($\phi$)—and employs 4 variational circuit layers. The model takes the Z-basis expectation values of the output of the first two qubits, which are scaled from [-1,+1] to [-$s$,+$s$] via a trainable parameter $s$ to enhance discriminative ability, and then processes them with a softmax function to generate probability distributions for cross-entropy classification. The QRU portion of this architecture includes 131 trainable parameters, and together with the scaling parameter $s$, the overall model has 132 trainable parameters. The full circuit design and a per-component parameter breakdown are presented in Supplementary Figure S5 and Section S3.4, respectively.

In the data preprocessing phase, we first standardize all pixel values, adjusting their range to [0,1]. Given QRU's sequential processing nature, we treat each 8×8 image as a sequence of length 8 with a feature dimension of 8. Specifically, each time step inputs one row of pixel data from the image, completing the processing of the entire image over 8 time steps.

The training process uses batch training with a batch size of 50. Although this experiment only uses the "3" and "5" digit classes, to maintain consistency with complete MNIST classification problems, we use cross-entropy rather than binary cross-entropy as the loss function:

$$loss = -\frac{1}{N}\sum_{i=1}^{N}\sum_{j=1}^{C} y_{ij}\log(\hat{y}_{ij})\,, \tag{9}$$

where $y_{ij}$ is the true label for the *i*-th sample in the *j*-th class (one-hot representation, taking values 0 or 1), $\hat{y}_{ij}$ is the model's predicted probability for the *i*-th sample in the *j*-th class, *N* is the number of samples, and *C* is the number of classes.

To prevent overfitting, an early stopping mechanism is implemented that records the validation set loss every 10 epochs. When no significant change is observed in 10 consecutive recorded validation losses (i.e., the last 9 loss values are not lower than the tenth-last loss value), or the loss value stabilizes with no further change, the training process is terminated.

### 3.4.2 Results Analysis and Comparison

In the MNIST digit recognition task for "3" and "5", the QRU model demonstrates stable, consistent performance. As shown in Table 6, in 7-fold stratified cross-validation, the QRU model achieved a mean training accuracy of 98.70% (standard deviation 0.30%) and a mean test accuracy of 98.05% (standard deviation 0.42%). Each fold maintained high performance, with training accuracy ranging from 98.17% to 99.11% and test accuracy ranging from 97.61% to 98.75%, demonstrating the model's stability in handling visual data. Training loss curves across all 7 folds are provided in Supplementary Figure S6, showing consistent convergence with no signs of training instability.

Table 6: QRU Model Performance in 7-fold Cross-validation

| | Training Accuracy | Testing Accuracy | Epochs |
|---|---|---|---|
| **Fold-1** | 98.67% | 97.76% | 410 |
| **Fold-2** | 98.17% | 97.76% | 970 |
| **Fold-3** | 98.88% | 98.75% | 810 |
| **Fold-4** | 98.91% | 98.39% | 740 |
| **Fold-5** | 98.60% | 97.87% | 690 |
| **Fold-6** | 98.56% | 98.23% | 440 |
| **Fold-7** | 99.11% | 97.61% | 840 |
| **Mean** | 98.70% | 98.05% | 700 |

Using only 132 parameters, QRU achieves 98.05% test accuracy on this task. In comparison, the benchmark testing by Bowles et al. [22] showed that a classical CNN required 51,841 parameters to achieve 96.42% accuracy. The same study also included a comparison with two quantum models: the Quanvolutional Neural Network (Quanvolutional NN) [45], which achieved 91.41% accuracy, and Wei Net [46], which achieved 70.49%. As shown in Table 7, QRU reduces parameter count by 99.7% compared to CNN while achieving higher accuracy. This benchmark test result is particularly valuable as it provides a direct comparison of classical models and different quantum approaches under the same conditions.

Table 7: Performance Comparison of Different Methods on MNIST "3" and "5" Classification

| Model | Parameters | Training Accuracy Mean (std) | Testing Accuracy Mean (std) |
|---|---|---|---|
| **QRU Model (This study)** | 132 | 98.70% (±0.30%) | 98.05% (±0.42%) |
| **CNN [22]** | 51,841 | 96.79% (±0.79%) | 96.42% (±0.64%) |
| **Quanvolutional NN [22,45]** | - | 93.24% (±1.10%) | 91.41% (±0.91%) |
| **Wei Net [22,46]** | - | 70.46% (±0.25%) | 70.49% (±0.25%) |

Values for CNN, Quanvolutional NN, and Wei Net are taken from the publicly released results of [22]; the CNN parameter count is obtained by instantiating the released model with its reported best hyperparameters for the 8×8 dataset.

The performance differences with quantum models in [22] may reflect fundamental design distinctions. Bowles et al. [22] observed that Quanvolutional NN's basis encoding degraded data quality through random feature mapping. Additionally, amplitude-encoding models (including WeiNet) consistently underperformed in their benchmarks, leading the authors to question whether "amplitude embedding [is] just not a good design choice" for their settings.

QRU's angle encoding with trainable scaling parameters and sequential C-SWAP-based information processing offers a contrasting approach that maintains information integrity while enabling adaptive feature selection throughout the temporal pipeline. These results validate QRU's capability in handling high-dimensional visual data with substantially fewer parameters than classical counterparts.

## 3.5 Preliminary Hardware Validation

### 3.5.1 Noise-Aware Training Strategy

Since real quantum hardware without quantum error correction implemented is inherently noisy, a quantum machine learning model intended for deployment on such devices must account for hardware noise during training. However, directly training on current NISQ devices is impractical: a single training run requires thousands of circuit execution iterations across many epochs, making QPU time and wall-clock time prohibitive due to limited accessibility and availability of QPU resources.

Two models for simulating noise that mimic real hardware noise are considered. First, a backend-derived noise model (abbreviated as NM) that incorporates gate errors, readout errors, and T1/T2-based thermal relaxation provides a partial alternative to real-device execution. Second, a fake-backend-based simulation model (e.g., provided by IBM Quantum) further accounts for hardware constraints such as qubit topology, coupling map, and basis gates. But both approaches can incur substantial computational cost and overhead for training simulations for quantum machine learning, as the qubit count and circuit depth increase.

We therefore adopt a noise injection (NI) approach — a noise-aware simulation-based training strategy — in which backend-specific noise is approximated by a lightweight transformation applied to statevector (SV) outputs, allowing the model parameters to adapt to simulated hardware noise characteristics prior to hardware evaluation. Note that the iterative NI training loop operates entirely within statevector simulation. The NM and fake backend are used to calibrate the noise-injection coefficients described below and to evaluate the trained models after training, but they are not directly invoked during each training iteration.

We model the effect of hardware noise on each observable expectation value $\langle O\rangle_{\text{noisy}}$ as:

$$\langle O\rangle_{\text{noisy}} = \alpha \cdot \langle O\rangle_{\text{ideal}} + \mathcal{N}\left(0, K^2\left(1-\alpha^2\langle O\rangle_{\text{ideal}}^2\right)\right), \tag{10}$$

where $\langle O\rangle_{\text{ideal}}$ is the expectation value of observable $O$ obtained from statevector simulation, $\mathcal{N}(\mu,\sigma^2)$ denotes a normal distribution with mean $\mu$ and variance $\sigma^2$, $\alpha \in (0,1]$ is a signal attenuation coefficient capturing the net reduction in observable magnitude due to gate noise accumulated over the quantum circuit, including contributions from depolarizing noise, dephasing, and related error sources, and $K > 0$ is a noise amplitude coefficient reflecting heteroscedastic shot noise, where the noise variance is not constant, but directly depends on the signal's underlying expected value.

To estimate the noise coefficients $\alpha$ and $K$ for a target backend, we designed an Observable-Targeted Calibration (OTC) procedure. For each single-qubit observable, a classical optimizer finds parameters $\vec{\theta}$ of a QRU parameterized quantum circuit such that the qubit output of the statevector expectation value $\langle O(\vec{\theta})\rangle_{\text{ideal}}$ matches a set of predefined target values spanning the observable range, e.g., (±0.8, ±0.4) — ensuring that the calibration uniformly covers the full dynamic range of the model's outputs. The QRU circuit with same parameters $\vec{\theta}$ is then executed on the target backend (e.g., the NM or FakeMarrakesh backend), yielding noisy expectation values at known ideal reference points. The pairs $\{(\langle O(\vec{\theta})\rangle_{\text{ideal}}, \langle O(\vec{\theta})\rangle_{\text{noisy}})\}$ are used to fit $\alpha$ and $K$ independently for each qubit and measurement basis (Z and X). Detailed calibration parameters for each backend and task are provided in Supplementary Section S5.

Using the calibrated coefficients $\alpha$ and $K$ for the target backend, noise is injected into the statevector outputs during training by applying the transformation in Eq. (10) to each observable at every time step. For the WDBC task, where the available dataset is limited, the training set is replicated five times; the stochasticity of noise injection ensures that identical inputs yield different outputs across replications. For the MNIST task, where the dataset is sufficiently large, no replication is applied.

The training configurations follow those described in Sections 3.3 and 3.4, with the following differences. First, the loss function is replaced by mean squared error (MSE) for both tasks, without the trainable scaling parameter used in the original experiments, with class labels correspondingly mapped from {0, 1} to {−1, +1} to match the circuit's native output range. Cross-entropy loss is known to perform well for classification tasks, but applying it directly to the circuit's inherently bounded [-1, +1] output caps the achievable classification confidence (sigmoid(±1) ≈ 0.73/0.27) regardless of how well-separated the classes are; this motivated the trainable scaling parameter in the original experiments, which amplified the output before the sigmoid to allow higher-confidence predictions. This modification reflects a hardware-motivated consideration, however: the trainable scaling parameter can allow near-zero circuit outputs to still yield strong classification performance under ideal statevector simulation, but such small differences are not physically meaningful once realistic hardware noise is present, as they fall within the noise floor. Removing the scaling parameter therefore returns the circuit output to its native bounded [-1, +1] range, for which cross-entropy is no longer well suited; MSE is adopted instead, as it regresses directly onto the {−1, +1} targets without requiring a sigmoid transform, and ensures that the learned separation remains meaningful in the presence of noise.

Second, for the WDBC task, LOOCV is replaced by a fixed train/validation/test split (340/113/116 samples), with dataset replication (×5) applied to the training and validation sets only, to reduce the cost of repeating NM and fake backend evaluation across all folds. Because noise injection introduces stochastic variation at each replication, this is effectively equivalent to training on five distinct noisy realizations of the dataset rather than five identical copies. Third, for the MNIST task, only the first fold of the seven-fold cross-validation is used to reduce this same evaluation cost across folds.

### 3.5.2 Noise Injection Results

Executions on the noise model (NM) and FakeMarrakesh backends used the default measurement shot count of 1024 per circuit evaluation.

Table 8: QRU Model Performance on WDBC Task under Noise Injection Training

| Training | SV test accuracy | Backend test accuracy |
|---|---|---|
| SV only (baseline) | 95.69% | 91.38% (NM) / 87.93% (Fake) |
| SV+NI (NM OTC) | 94.83% | 94.83% (NM) |
| SV+NI (Fake OTC) | 91.38% | 88.79% (Fake) |

Table 8 summarizes the classification test accuracy of QRU models trained under different noise conditions and evaluated on their respective target backends for the WDBC task. Without NI, the SV-trained baseline achieves 95.69% under ideal simulation but degrades to 91.38% and 87.93% on the noise model (NM) and fake backend (FakeMarrakesh), respectively. NI training with NM OTC calibration recovers this degradation, achieving 94.83% on the noise model. NI training with Fake OTC calibration similarly maintains 88.79% on the fake backend.

Table 9: QRU Model Performance on MNIST Task under Noise Injection Training

| Training | SV test accuracy | Backend test accuracy |
|---|---|---|
| SV only (baseline) | 97.09% | 49.74% (NM) / 51.40% (Fake) |
| SV+NI (NM OTC) | 85.80% | 84.65% (NM) |
| SV+NI (Fake OTC) | 83.82% | 77.06% (Fake) |

Table 9 shows the corresponding results for the MNIST task. Without NI, the SV-trained baseline achieves test accuracy of 97.09% under ideal simulation but degrades sharply to 49.74% and 51.40% on the NM and FakeMarrakesh, respectively — near random-chance performance for a binary classification task. NI training with NM OTC calibration recovers performance to 84.65% on the NM, and NI training with Fake OTC calibration achieves 77.06% on the fake backend (FakeMarrakesh). These results, together with those of the WDBC task, suggest that noise-aware simulation-based training is a viable strategy for adapting quantum models to NISQ hardware noise characteristics.

### 3.5.3 Error Mitigation on Real Hardware

To assess the feasibility of running QRU on real quantum hardware, we employed QESEM (Quantum Error Suppression and Error Mitigation), a commercial error mitigation service by Qedma available as an IBM Qiskit Function [47], which combines error suppression and unbiased quasi-probabilistic error mitigation to achieve high-accuracy expectation value estimation on noisy QPUs [48]. All real hardware experiments presented for Table 10 and Figure 7 were conducted on ibm_marrakesh, except for probabilistic error cancellation (PEC) [49, 50]. The circuit sampling overhead on ibm_marrakesh was too large for the available QPU time, so PEC should be measured on another QPU with much less circuit sampling overhead, which grows exponentially with the circuit's total noise. PEC was therefore measured on ibm_phoenix, a Nighthawk r2 QPU that became available after the other measurements in this section were completed and on which this circuit requires fewer CZ gates (Supplementary Section S8). These experiments and the simulation results (NM and FakeMarrakesh) in this section used approximately 4000 shots per circuit evaluation, except for QESEM and PEC, which determine their shot budgets internally from the requested precision and from the learned sampling overhead, respectively. Single-step error mitigation experiments (Table 10, Figure 7) were configured with a QESEM precision of 0.02. Sequential inference experiments (Figures 8 and 9) were configured with a QESEM precision of 0.05.

Table 10: Error Mitigation Comparison (Single-Step Circuit)

| Method | Mean deviation from SV |
|---|---|
| **Simulation** | |
| Statevector (reference) | - |
| Noise model (ibm_marrakesh) | −13.0% |
| FakeMarrakesh | −38.1% |
| **Real hardware** | |
| Readout mitigation only (ibm_marrakesh) | −78.2% |
| Twirling (300 inst., ibm_marrakesh) | −83.4% |
| ZNE (32 twirling inst., ibm_marrakesh) | −70.9% |
| ZNE (300 twirling inst., ibm_marrakesh) | −65.2% |
| PEC (15 runs, ibm_phoenix) | −6.0% |
| QESEM (ibm_marrakesh) | +0.3% |

Table 10 compares the performance of different error mitigation strategies using a single-step QRU circuit from the MNIST task, with statevector simulation serving as the reference; each deviation is the mean of the signed relative deviations (method − SV)/SV over the five observables. All real-hardware Qiskit Runtime rows use the default readout mitigation (twirled readout error extinction, TREX). The twirling and both zero-noise extrapolation (ZNE) rows additionally apply Pauli gate twirling, averaged over the indicated number of randomly twirled circuit instances (inst.), and ZNE uses noise scale factors of 1, 3, and 5. PEC, also available in Qiskit Runtime, was applied with noise learning on all distinct layers of entangling gates and without a limit on its sampling overhead. Its deviation is the mean over 15 independent runs with the number of circuit randomizations set to 400–1,600 (Supplementary Section S8). The NM and FakeMarrakesh rows are included as simulated-noise baselines without any error mitigation applied, shown here to contextualize how closely these simplified simulations approximate real hardware behavior.

The discrepancy between the simulated-noise baselines and the readout-mitigated real-hardware result may be attributed to several factors associated with the default simulation settings and the simplified nature of backend-derived noise models. First, in the NM simulation used here, the circuit was simulated with a backend-derived noise model, but the target backend's coupling map was not explicitly enforced during transpilation in the same way as in the FakeMarrakesh simulation [51]. As a result, the NM simulation does not fully account for the routing overhead imposed by the hardware topology. In contrast, the FakeMarrakesh simulation enforces the target device's native coupling map and basis gates, requiring additional routing operations to satisfy device connectivity constraints and thereby increasing the accumulated noise. Standard backend-derived noise models, based on static backend calibration data, also do not fully capture complex hardware noise mechanisms such as crosstalk, correlated or nonlocal errors, and non-Markovian noise dynamics [52, 53]. Second, because the simulations used here do not include explicit scheduling and delay instructions, idle-time relaxation during periods when other qubits are operated on may not be fully represented [51]. The NM and FakeMarrakesh rows should therefore be interpreted as simplified simulated-noise baselines rather than quantitatively accurate predictors of real-hardware behavior.

We additionally tested this interpretation using a more hardware-aware single-step simulation configuration that more closely matches real-device execution settings (Supplementary Section S6). Under this configuration, where qubit topology, coupling map, native basis gates, and gate durations are explicitly enforced during transpilation with an explicit time scheduling pass for both the NM and FakeMarrakesh models, the two models therefore converge more closely to each other, and the deviations from the statevector references become −59.7% and −59.8%, respectively, substantially closer to the readout-mitigated real-hardware deviation of −78.2%. The residual discrepancy relative to the real-hardware result may be due to the simplified backend-derived noise models not fully capturing more complex physical noise processes, including crosstalk, correlated or nonlocal error channels, non-Markovian noise dynamics, etc. [52, 53]. This more hardware-aware configuration was not used for the noise-injection training experiments in Sections 3.5.1–3.5.2, which require repeated simulation across many training iterations and therefore adopt lighter default settings for computational feasibility. The NI-trained classification accuracies reported in Tables 8 and 9 (88.79% for WDBC, 77.06% for MNIST) are therefore calibrated and evaluated only against these simplified, default-setting simulations; whether comparable performance would transfer to real-hardware execution, where the deviation is substantially larger even with readout mitigation, remains unverified.

The observables are evaluated at circuit parameters selected via OTC such that their statevector expectation values are approximately +0.8. With Qiskit Runtime's default readout mitigation, the real hardware result deviates from the statevector reference by −78.2%, and gate twirling and ZNE also fail to recover meaningful agreement, with deviations ranging from −65.2% to −83.4%. PEC gives a mean deviation of −6.0% over 15 independent runs (range −3.2% to −9.7%), with all runs below the reference. On ibm_phoenix, readout mitigation and gate twirling applied to the circuit used for PEC deviate by −43.8% and −38.0%, and ZNE by +24.9% and +29.8% (Supplementary Section S8). QESEM recovers the statevector reference to within 0.3% at a requested precision of 0.02. At this operating point and with the tested settings, high-fidelity expectation value estimation for this QRU circuit is therefore achievable on current

NISQ hardware using QESEM. The QESEM deviation over the range of expectation values visited during inference is examined below with the data in Figure 8.

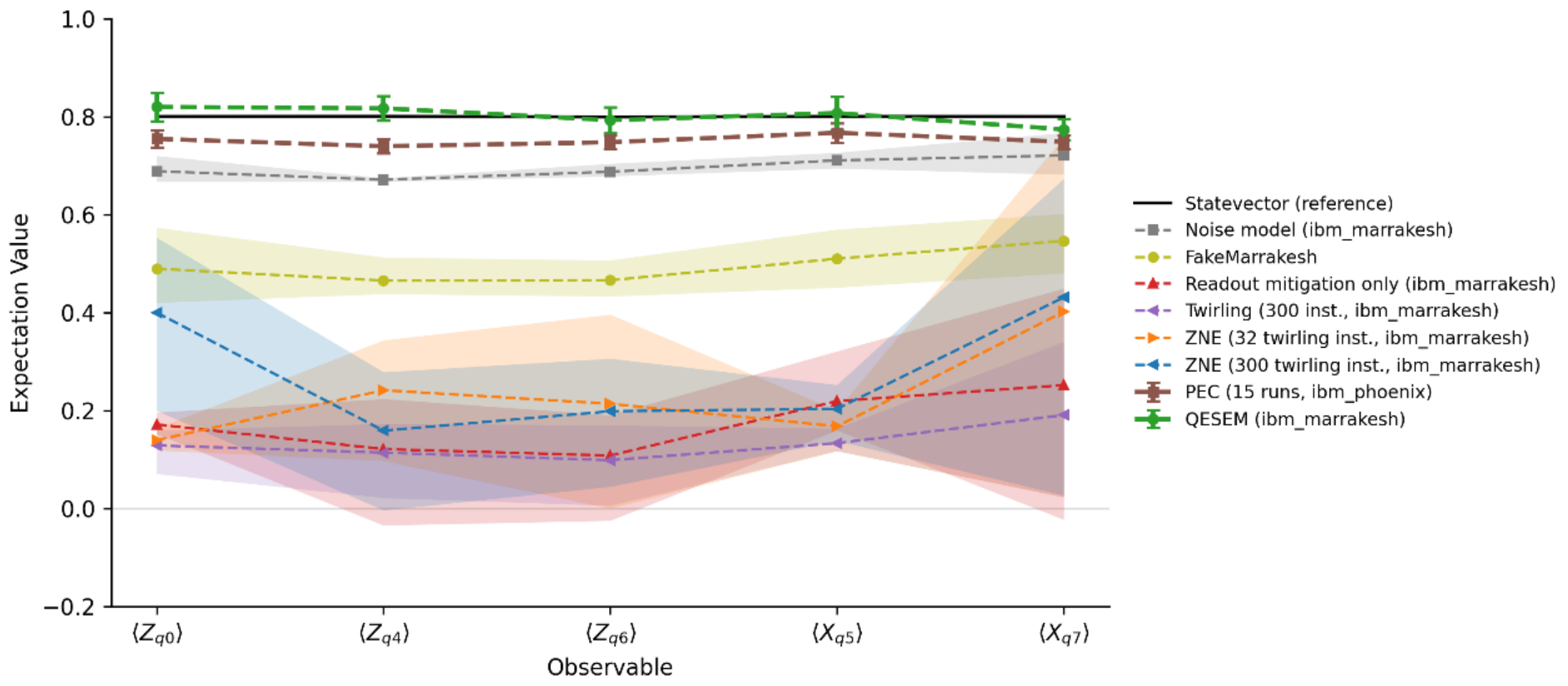

Figure 7: Comparison of error mitigation methods across five single-qubit observables of a single-step MNIST QRU circuit. Shaded regions indicate the range between minimum and maximum values across three repeated runs; QESEM is shown as a point estimate with its reported uncertainty, and PEC as the mean over 15 runs with the run-to-run standard deviation. Method labels are as in Table 10, with settings described in Section 3.5.3; "inst." denotes the number of randomly twirled circuit instances.

Figure 7 shows the individual observable results underlying the averages reported in Table 10. QESEM recovers expectation values close to the statevector reference across all five observables, PEC lies 4.0%–8.0% below it for every observable, and all other methods remain substantially below the reference regardless of the specific observable measured.

To further assess the feasibility of sequential inference on real hardware, we ran complete 8-step MNIST classification using QESEM on two IBM Heron-generation backends: ibm_marrakesh (Heron r2) and ibm_boston (Heron r3), each on an independent test sample (a digit-3 sample on ibm_marrakesh and a digit-5 sample on ibm_boston). The QESEM measurements shown in Figures 8 and 9 are obtained from the same real-hardware sequential inference runs; the two figures differ only in how the statevector (SV) reference is constructed.

For Figure 8, the z-score, defined as (QESEM − SV) / σ, with σ denoting the QESEM uncertainty, at each step is computed using SV expectation values that are recomputed using the hidden state vector recorded from the actual QESEM circuit execution at that step. This provides a stepwise conditional comparison that isolates the mitigation quality of each individual circuit execution under the same hidden state input, independent of the accumulated mismatch between the QESEM and independently propagated SV trajectories from earlier steps. In Figure 9, by contrast, the SV reference is independently propagated sequentially across the full 8-step sequence, without using QESEM intermediate hidden states. At each step, we compute the L2 distance $\|\vec{h}_{QESEM} - \vec{h}_{SV}\|$ between the QESEM hidden state vector obtained from the real-hardware run $\vec{h}_{QESEM}$ and the corresponding independently propagated SV hidden state reference $\vec{h}_{SV}$; the two curves in Figure 9 correspond to the two backend/sample pairs, namely ibm_marrakesh for the digit-3 sample and ibm_boston for the digit-5 sample.

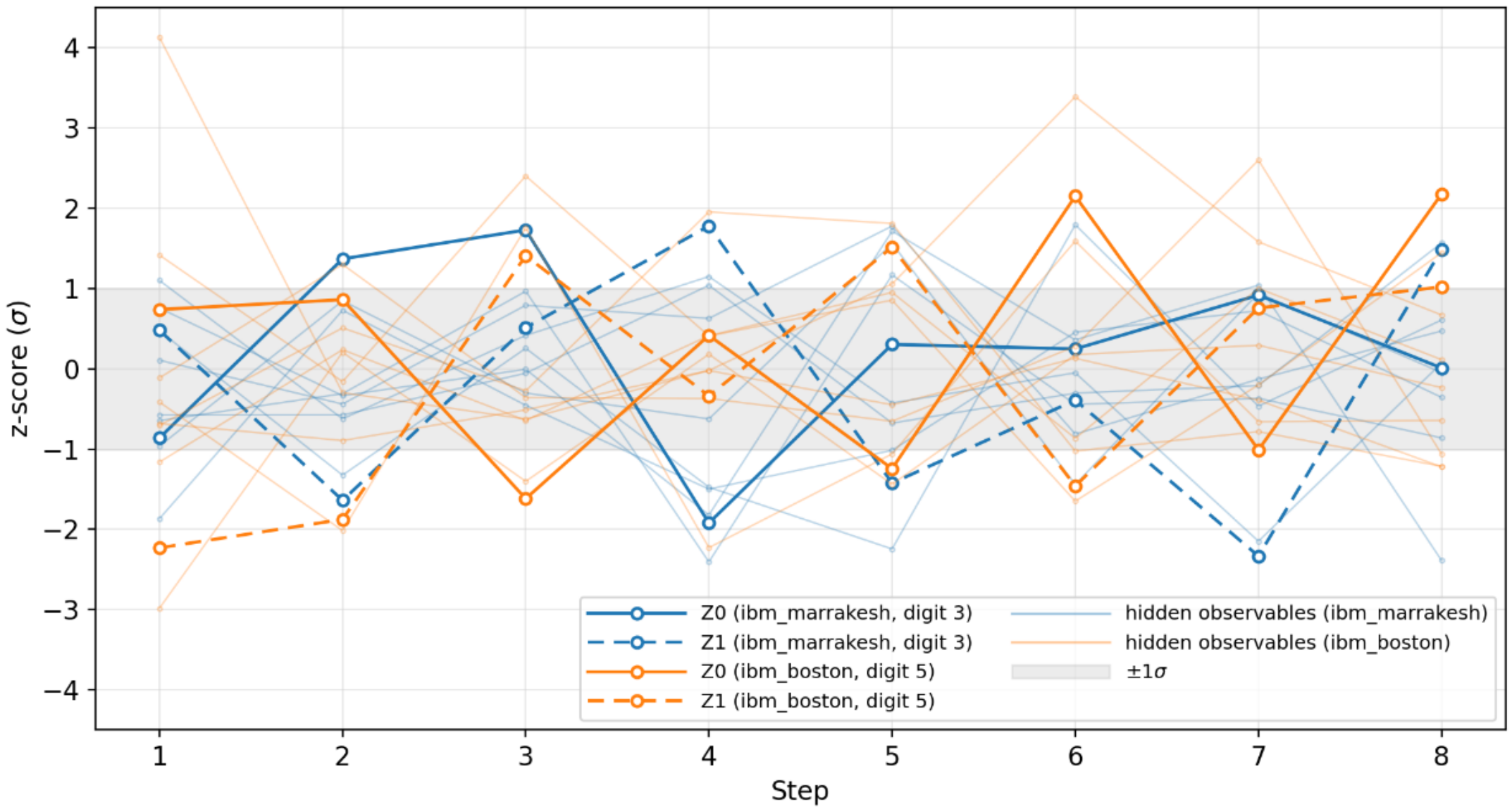

Figure 8: Stepwise z-score comparison between QESEM mitigated and SV expectation values for two independent MNIST inference samples. At each step, the SV reference uses the same hidden state input as the corresponding QESEM circuit, isolating per-step mitigation quality under matched input conditions. Output observables ($Z_0$, $Z_1$) are highlighted for each backend; hidden state observables are shown as faint lines. The shaded band indicates ±1σ.

Figure 8 shows the per-step z-score at each observable. Because the sequential inference visits a wide range of expectation values, these data also characterize the QESEM deviation beyond the single OTC operating point of Table 10. Across all 160 recorded observables (2 backends × 8 steps × 10 observables), whose statevector values span [−0.79, +0.75], the median absolute deviation from the statevector reference is 0.018 (maximum 0.134), and 59% and 91% of the points lie within ±1σ and ±2σ, respectively, compared with 68% and 95% expected if the deviations were purely statistical and fully captured by σ for a normal distribution. The deviation shows a small systematic shrinkage toward zero, corresponding to recovery of approximately 95% of the ideal expectation-value magnitude at the precision used for sequential inference (0.05).

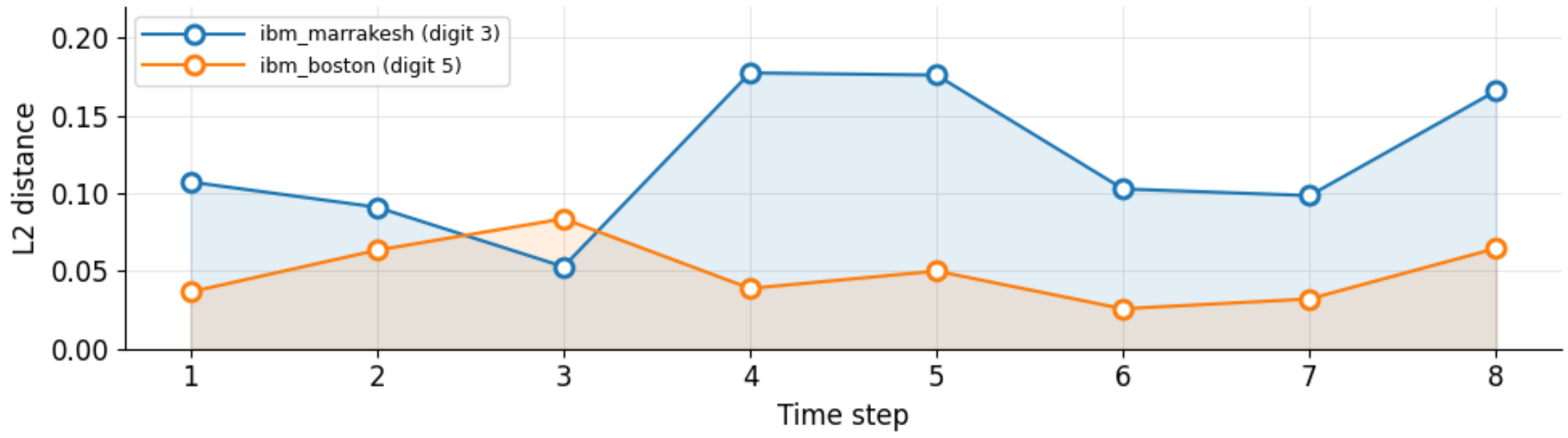

Figure 9: L2 distance between the QESEM hidden state vector and an independently propagated SV reference at each time step, for two MNIST inference samples on ibm_marrakesh and ibm_boston. Each curve corresponds to one backend/sample pair, not to QESEM or SV separately. No systematic error accumulation is observed, and both backend/sample cases yield the correct final classification in both QESEM and SV inference.

Figure 9 shows the L2 distance $\left\|\vec{h}_{QESEM} - \vec{h}_{SV}\right\|$ between the QESEM hidden state vector and the independently propagated SV reference at each time step. Both runs exhibit step-to-step fluctuation, but no systematic error accumulation across the sequence, and both SV and QESEM runs yield the correct final classification on both backends. The ibm_boston run shows a smaller overall L2 distance than ibm_marrakesh; however, with only one sample per backend, we cannot determine whether this reflects a hardware-specific effect.

Sequential QESEM inference is substantially more costly than the single-step comparison in Table 10 and Figure 7, both in QPU usage and, more importantly, in wall-clock time. The QESEM protocol requires additional characterization and mitigation circuit executions for each requested observable set [47, 48], so a complete sequential inference run involves far more circuit jobs than the number of QRU time steps alone. In our IBM Quantum job records for one complete 8-step run, the execution required 33 completed batch jobs, each internally comprising several dozen circuit-execution sub-jobs, and totaling 4301 seconds of reported QPU usage, corresponding to approximately 1.2 hours. Moreover, because the input to each QRU step depends on the hidden state obtained from the previous step, the eight time steps cannot be launched in parallel. Including queueing, service-side mitigation, and post-processing overhead, one complete 8-step sequential inference run required approximately one week of wall-clock time under our execution conditions. Given this cost, sequential inference validation in this work is limited to one representative sample per backend, and should be interpreted as a proof-of-concept hardware feasibility test rather than a statistical performance evaluation.

# 4. Comprehensive Analysis and Discussion

## 4.1 Efficiency Analysis

This study systematically validates QRU's efficiency advantages in parameter usage and quantum resource allocation through three different types of experiments. Table 11 summarizes the parameter usage and performance comparison between QRU and other models across different task types, based on statevector simulation results from Sections 3.2–3.4.

Table 11: Cross-task Parameter Efficiency Comparison. Parameter reduction is computed as ((Model Parameters − QRU Parameters) / Model Parameters) × 100%.

| Task Type | Model | Parameters | Test Performance | Parameter Reduction |
|---|---|---|---|---|
| **Time Series Prediction** | QRU | 72 | MSE: 1.94E-02 | Baseline |
| | GRU(77) | 77 | MSE: 1.60E-01 | 6.5% |
| | GRU(197) | 197 | MSE: 7.28E-02 | 63.5% |
| | GRU(307) | 307 | MSE: 1.80E-02 | 76.5% |
| | LSTM | 491 | MSE: 4.69E-01 | 85.3% |
| **WDBC Classification** | QRU | 35 | ACC: 96.13% | Baseline |
| | MLP, 2 hidden neurons | 65 | ACC: 96.84% | 46.2% |
| | MLP, 14 hidden neurons | 449 | ACC: 97.36% | 92.2% |
| | MLP, WEKA default | 530 | ACC: 97.01% | 93.4% |
| | MLP [40] | 65–449† | ACC: 96.21% | 86.4%*† |
| | ANN [41] | 167 | ACC: 96.50% | 79.0% |
| | DNN [42] | 1,601† | ACC: 95.32% | 97.8%† |
| **MNIST Recognition** | QRU | 132 | ACC: 98.05% | Baseline |
| | CNN [22] | 51,841 | ACC: 96.42% | 99.7% |

*Calculated using the median of the MLP parameter range.
†Estimated (see Table 5); reductions marked † inherit the estimate.

QRU demonstrates consistent parameter efficiency advantages under statevector simulation. In time-series prediction tasks, a 72-parameter QRU achieves performance comparable to that of a 307-parameter GRU (full results) or a 197-parameter GRU (after outlier removal), representing a 76.5% or 63.5% parameter reduction, respectively. In WDBC classification tasks, QRU using 35 parameters achieves 96.13% test accuracy, comparable to MLPs trained in this study under the same protocol with 65–530 parameters (96.84%–97.36%), and to ANN requiring 167 parameters (96.50%) and DNN requiring 1,601 parameters (95.32%) reported in the literature. In MNIST digit recognition tasks, QRU achieves higher accuracy (98.05%) with 132 parameters compared to CNN's 96.42% accuracy using 51,841 parameters. Overall, QRU achieves 46.2%–99.7% parameter reduction compared to classical counterparts across all three tasks.

The significant reduction in parameter count demonstrated in our experiments can be attributed to QRU's architectural efficiency. While classical GRU architecture requires separate neural networks with substantial parameters for each component (update gate, reset gate, and candidate generation), QRU achieves similar functionality through quantum-native operations—specifically, using C-SWAP gates for information selection, which eliminates the need for multiple parameter-heavy networks.

Moreover, QRU's efficient parameter scaling is maintained regardless of input dimension, as demonstrated in our experiments across different tasks. This consistent efficiency makes QRU particularly attractive for large-scale applications where parameter efficiency is critical. While other quantum approaches might achieve similar parameter reductions, they often require deeper circuits or multiple VQCs, therefore limiting their practicality in NISQ devices [22, 45, 46, 54–56].

Regarding quantum resource allocation, QRU achieves unique efficiency through its recurrent structure. This design allows sequential processing of high-dimensional input features through limited qubits while effectively retaining and integrating key information. Across the three experiments, QRU used 8, 7, and 10 qubits, respectively, including input, hidden state, and ancilla qubits. This stable quantum resource usage, combined with the structured variational circuit design, demonstrates QRU's practical efficiency under NISQ constraints, where both qubit count and circuit depth are critical considerations. Importantly, each QRU terminates with quantum measurements and feeds the measurement results forward to the next time step, ensuring that the per-step quantum circuit depth remains effectively confined within a single QRU regardless of sequence length, thereby mitigating the stringent quantum circuit-depth constraints of NISQ devices.

## 4.2 Model Characteristics Analysis

Through three progressive experiments, we observe several key characteristics of QRU that demonstrate its potential beyond parameter efficiency. These characteristics stem from the integration of recurrent structure and C-SWAP operations, as well as the measurement results feedforward scheme, showing how these quantum-native mechanisms can be effectively utilized for neural network design under NISQ constraints.

Unlike standard VQC approaches that often require fixed input-output dimensions, QRU's recurrent structure enables processing of arbitrary-dimensional data in a sequential manner. Our experiments across different domains - from time-series prediction to image classification - validate this flexibility, demonstrating that the architecture can effectively handle diverse data types and tasks.

The information update mechanism in QRU leverages quantum-native C-SWAP operations, representing an approach distinct from classical neural network design principles. Rather than relying on parametric transformations to select and mix previous and present information, QRU implements this information selection mechanism via C-SWAP operations (see Section 2.2.3). This design demonstrates how such quantum operations enable efficient neural information processing, particularly in implementing state update and reset operations with minimal quantum resources.

The integration of quantum information selection and recurrent processing with the measurement results feedforward scheme demonstrates QRU's potential as a building block for modern machine learning architectures. This design approach suggests new possibilities for developing quantum neural network components that are both versatile and hardware-aware, particularly valuable in the NISQ era. The success of this implementation opens the door to integration into various machine learning architectures, providing a practical template for future quantum machine learning designs.

## 4.3 Limitations and Future Prospects

Several practical limitations should be acknowledged. First, circuit compilation onto real hardware introduces substantial overhead: e.g., a single C-SWAP gate expands to approximately 53 native gates (11 CZ, 18 SX, 24 RZ) on current IBM devices (at optimization level 1, ibm_marrakesh), resulting in the total per-step transpiled quantum circuit depths of approximately 250 and 420 for the WDBC and MNIST configurations, respectively, corresponding to a factor of approximately 7–9× relative to the abstract-level depth. Second, QRU's sequential processing structure requires fully executing and measuring each step's circuit before the next step begins, meaning that wall-clock execution time scales linearly with sequence length and cannot be parallelized. Combined with current QPU queue times, this makes hardware training impractical at present; therefore, the performance results reported in Sections 3.2–3.4 are based on statevector simulation. Third, the noise from measurement shots of real hardware introduces additional uncertainty, and error mitigation strategies such as QESEM and PEC, which gave mean deviations of +0.3% and −6.0%, respectively, in the settings tested in Section 3.5.3, require substantial QPU resources that limit their scalability. The hardware experiments presented in Section 3.5 should therefore be understood as a preliminary proof of concept rather than a full validation of QRU's performance on real devices.

Trainability is a further general concern for variational quantum circuits, where barren plateaus (BPs) can impede optimization in large-qubit-count, deep-circuit regimes [57, 58]. However, the per-step circuits in this study involve at most 10 qubits and have a fixed, relatively shallow circuit depth, placing them in a low-risk BP regime. Training converged consistently across all three tasks with no signs of gradient vanishing (Supplementary Figures S2, S4, and S6). In addition, QRU's C-SWAP mechanism inherently injects fresh $|0\rangle$ states at each time step, which may constrain the explored quantum state space; whether this has any mitigating effect on BP tendencies — and more broadly, whether the low-risk regime observed here persists as qubit count or circuit depth increases — remains an open theoretical question. A related consideration concerns training directly on real quantum hardware. Because gradients cannot be obtained via backpropagation on physical devices, hardware-based training would require the parameter-shift rule. Under a straightforward parameter-shift implementation, the reuse of shared parameters across multiple time steps would increase the number of circuit evaluations needed to estimate hardware gradients as the sequence length grows. This cost is not addressed by the present statevector-based training and would need to be considered for future hardware-based training. Stochastic gradient estimation methods that do not require separate parameter-shift evaluations for every trainable parameter, such as Simultaneous Perturbation Stochastic Approximation (SPSA) [59, 60], could reduce the gradient-estimation component of this cost, although the cost of each circuit evaluation would still increase with sequence length. Layer-wise learning [61], identity initialization [62], or restricted entanglement strategies [57] could mitigate barren plateaus should they arise at larger scale.

On the theoretical side, a rigorous framework for analyzing quantum state evolution within QRU and its relationship to parameter efficiency would strengthen the foundation of the architecture. Whether the C-SWAP-based $|0\rangle$ injection mechanism has any mitigating effect on BP tendencies also warrants further theoretical investigation. From an architectural perspective, exploring QRU's integration into larger machine learning systems, such as Transformer-based models, and evaluating its scalability on larger datasets and longer sequences are natural next steps. On the hardware side, extending the preliminary validation presented in Section 3.5 to more complex tasks and larger qubit configurations, as well as reducing compilation overhead through hardware-aware circuit design, would bring QRU closer to practical deployment on NISQ devices. Finally, as discussed in Section 2.2.6, a fully quantum recurrent architecture that bypasses inter-step measurement and re-encoding entirely remains a longer-term goal contingent on advances in fault-tolerant quantum computing.

# 5. Conclusion

The QRU model presented in this study represents a significant step toward quantum recurrent neural network architectures for the current NISQ era. By carefully designing quantum circuits that respect hardware constraints while leveraging quantum-native operations for efficient information processing, we have created an architecture that achieves consistent parameter efficiency across diverse applications under statevector simulation. Through three different types of experiments, we systematically validated QRU's performance and versatility: from basic time-series prediction to high-dimensional feature classification in medical diagnosis, and finally to visual data recognition. Across all experimental tasks, QRU achieves performance comparable to the classical approaches while consistently requiring fewer parameters. Separately, preliminary hardware inference experiments on IBM quantum processors provide initial evidence of QRU's execution feasibility on real NISQ devices.

These results support the promise of building quantum machine learning systems from the ground up, using quantum operations to implement neural network functionality rather than simply translating classical architectures into quantum circuits. The controlled-SWAP-based information selection mechanism, measurement results feedforward recurrent processing, and shared parameters across time steps enable efficient parameter utilization while maintaining the essential temporal processing capabilities of recurrent networks and provide a template for developing other quantum neural network components.

As quantum hardware continues to improve with increasing qubit counts, longer coherence times, and higher gate fidelities, architectures like QRU may become increasingly practical and powerful. The principles demonstrated here- quantum-native design, NISQ compatibility, parameter efficiency, and architectural versatility- point toward a future where QRU and similar architectures could serve as quantum-native building blocks within larger machine learning systems, offering a candidate quantum-native alternative to classical neural network components as hardware matures.

Ultimately, this work contributes to the ongoing effort to harness quantum mechanical properties for practical computational advantages. By demonstrating that QRUs can achieve performance comparable to classical neural networks with fewer parameters under statevector simulation, while maintaining a constant per-step effective quantum circuit depth regardless of input sequence length, we provide evidence that this approach represents a promising direction for quantum machine learning research.

## Data Availability

The code used to produce the results presented in this work is publicly available at https://github.com/lostsodium/quantum-recurrent-unit.

## Funding

H.-S. Goan acknowledges support from the National Science and Technology Council (NSTC), Taiwan, under Grants No. NSTC 113-2112-M-002-022-MY3, No. NSTC 113-2119-M-002-021, No. NSTC 114-2119-M-002-018, No. NSTC 114-2119-M-002-017-MY3, No. NSTC 115-2119-M-002-005 and No. NSTC 115-2927-I-002-535 and from the National Taiwan University under Grants No. NTU-CC-115L8937, No. NTU-CC-115L893704 and No. NTU-CC-115L8512. H.-S. Goan is also grateful for the support of the "Center for Advanced Computing and Imaging in Biomedicine (NTU-115L900702)" through the Featured Areas Research Center Program within the framework of the Higher Education Sprout Project by the Ministry of Education (MOE), Taiwan, the support of Taiwan Semiconductor Research Institute (TSRI) through the Joint Developed Project (JDP) and the support from the Physics Division, National Center for Theoretical Sciences, Taiwan.

## Author Contributions

T.-D.W. conceived the QRU architecture, implemented the quantum circuits, performed all experiments, analyzed the results, and wrote the manuscript. H.-S.G. conceived the project, supervised the research, provided critical guidance on methodology and interpretation, secured funding, and revised the manuscript. Both authors reviewed and approved the final manuscript.

## Competing Interests

Author Tzong-Daw Wu declares no competing interests. Author Hsi-Sheng Goan is an Associate Editor of EPJ: Quantum Technology, International Journal of Quantum Information, and Chinese Journal of Physics. Hsi-Sheng Goan was not involved in the journal's review of, or decisions related to, this manuscript.

## Additional Information

**Supplementary information:** Supplementary information is provided in a separate file.

# Supplementary Material for "Quantum Recurrent Unit: An Effective and Parameter-Efficient Quantum Neural Network Architecture for NISQ Devices"

Tzong-Daw Wu [1] and Hsi-Sheng Goan [1,2,3,*]

[1]*Department of Physics and Center for Theoretical Physics, National Taiwan University, Taipei 106319, Taiwan*

[2]*Center for Quantum Science and Engineering, National Taiwan University, Taipei 106319, Taiwan*

[3]*Physics Division, National Center for Theoretical Sciences, Taipei 106319, Taiwan*

* Email: goan@phys.ntu.edu.tw

# Supplementary Material: Detailed QRU Implementation

This document provides comprehensive implementation details for the three experimental validations of the Quantum Recurrent Unit (QRU) architecture described in the main text, as well as ablation studies conducted to validate key architectural design choices. Sections S1–S3 include detailed circuit specifications, parameter configurations, and processing methodologies for the oscillatory behavior prediction, WDBC classification, and MNIST digit recognition experiments, complementing the conceptual framework presented in Section 2 of the main text. Section S4 presents ablation experiments examining the C-SWAP information selection mechanism, the dual-basis measurement strategy, and the robustness of QRU's performance to input feature ordering.

## S1. Oscillatory Behavior Prediction Implementation

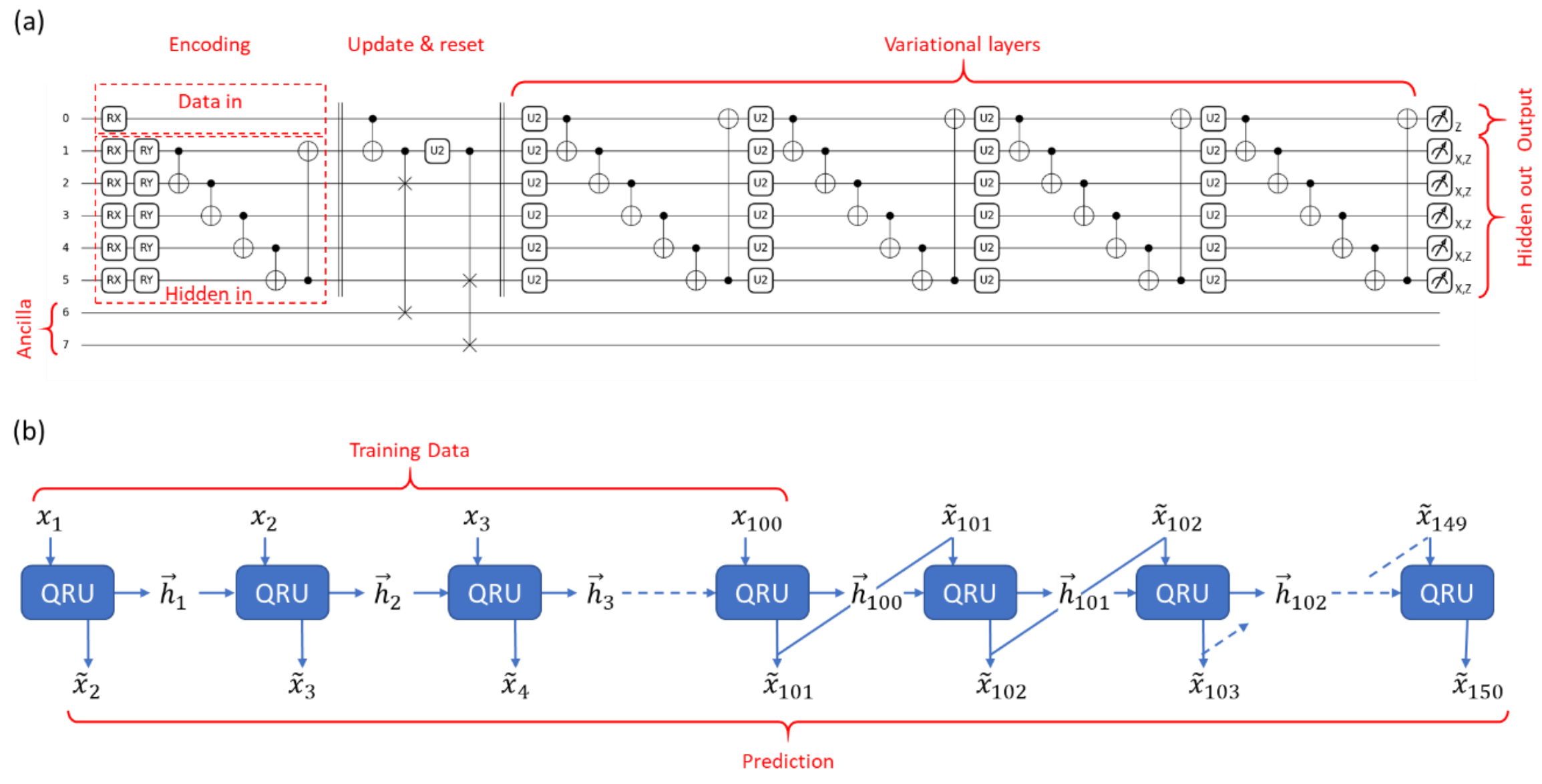

Figure S1: QRU circuit implementation for oscillatory behavior predictions. (a) Circuit architecture with 1 data qubit and 5 hidden qubits. (b) Sequential processing flow showing training and prediction phases.

### S1.1 Circuit Architecture

The oscillatory prediction experiment employs a QRU configuration explicitly designed for one-dimensional time series modeling (Figure S1(a)). The circuit uses 1 data qubit (qubit 0) for input encoding and 5 hidden qubits (qubits 1–5) for temporal state representation, providing a 10-dimensional hidden state space through Z and X basis measurements.

Input encoding utilizes a single Rx rotation gate with trainable parameters suitable for one-dimensional scalar input. The choice of Rx gates provides sufficient expressivity for encoding oscillatory patterns while maintaining circuit simplicity.

Hidden state encoding follows the same angle encoding scheme as described in Figure 2(b) of the main text. The 10-dimensional hidden state vector (obtained from measuring 5 qubits in both Z and X bases at the previous time step) is encoded into the 5 hidden qubits using Rx and Ry rotation gates with trainable parameters. Each qubit encodes two dimensions of hidden state information through its Rx and Ry rotations, with circular CNOT entanglement operations implementing correlations between qubits. Each encoding rotation in input data encoding and hidden-state encoding carries two trainable parameters, a weight and a bias ($\theta = w \cdot x + b$, see Eq. (5) in the main text).

The update & reset mechanism implements controlled-SWAP operations on selected qubits to enable information-selective processing, mimicking the classical GRU's update gate functionality. Not all qubits participate in swap operations, allowing partial retention of information that balances the integration/mixing of old and new information.

Four variational layers form the core computational component, each containing U2($\varphi$, $\lambda$) rotation gates with 2 trainable parameters per gate. U2 gates are two-parameter single-qubit rotations implementing Rz($\varphi$)Ry($\pi$/2)Rz($\lambda$), providing rich expressivity for function approximation while maintaining compatibility with NISQ device constraints.

## S1.2 Sequential Processing Framework

The experiment processes two types of oscillatory data through the sequential framework shown in Figure S1(b). The processing follows a step-by-step prediction paradigm where the model receives a single scalar input $x_i$ and predicts the next time step value $\tilde{x}_{i+1}$, with hidden states $\vec{h}_i$ propagated between steps.

The training phase (steps 1–100) uses true data values ($x_1$– $x_{100}$) from both oscillation types, enabling the model to learn the underlying dynamics of both patterns simultaneously. This joint training strategy tests the model's capability to distinguish and jointly learn different oscillatory behaviors within the same trained model.

The prediction phase ($\tilde{x}_{101}$– $\tilde{x}_{150}$) implements a self-feedback mechanism where each prediction output becomes the input for the subsequent time step. Specifically, the output $\tilde{x}_{101}$ from step 100 serves as input for step 101, and this pattern continues through step 149, producing final prediction $\tilde{x}_{150}$. This self-feedback approach evaluates the model's ability to maintain long-term temporal dependencies and generate coherent predictions without external ground truth.

## S1.3 Training Convergence

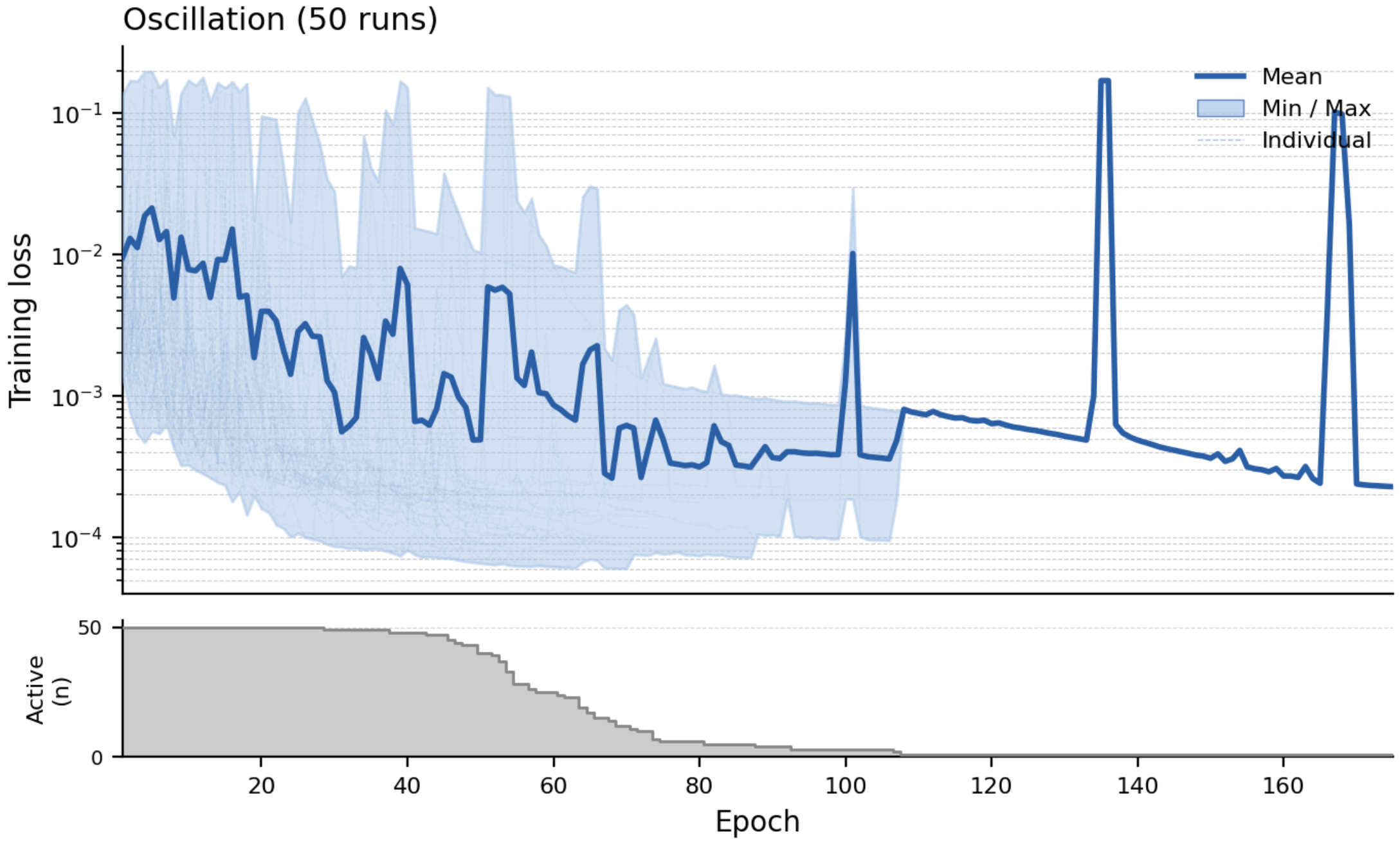

Figure S2: Training loss curves for the oscillation prediction task across 50 independent runs. Upper panel: mean training loss (solid line) with min/max range (shaded region) and individual runs (dashed lines), plotted on a logarithmic scale. Lower panel: number of active runs remaining at each epoch (early stopping applied individually per run).

Training convergence was evaluated across 50 independent runs with random parameter initializations. As shown in Figure S2, the mean training loss decreases consistently from approximately $10^{-1}$ at epoch 1 to approximately $2 \times 10^{-4}$ by the final epoch, demonstrating stable optimization throughout training. Occasional spikes in the mean loss curve (visible near epochs 105, 135, and 170) reflect the early termination of lower-loss runs, which shifts the mean upward transiently before it resumes its downward trend. The progressive reduction in active run count (lower panel) confirms that early stopping is triggered at varying epochs across runs, consistent with the diversity of convergence trajectories observed in the upper panel. No evidence of barren plateaus or systematic trainability failure is observed across the 50 runs.

## S1.4 Technical Specifications

- Total qubits: 8 (1 data + 5 hidden + 2 ancilla)
- Total trainable parameters: 72 (input encoding 2; hidden-state encoding 20; update-and-reset rotation 2; variational layers 48)
- Input dimension: 1 (scalar time series values)
- Output dimension: 1 (scalar prediction)
- Hidden state dimension: 10 (5 qubits × 2 bases)

- Training data: 100 time steps per oscillation type

- Test data: 50 prediction steps with self-feedback

## S2. WDBC Classification Implementation

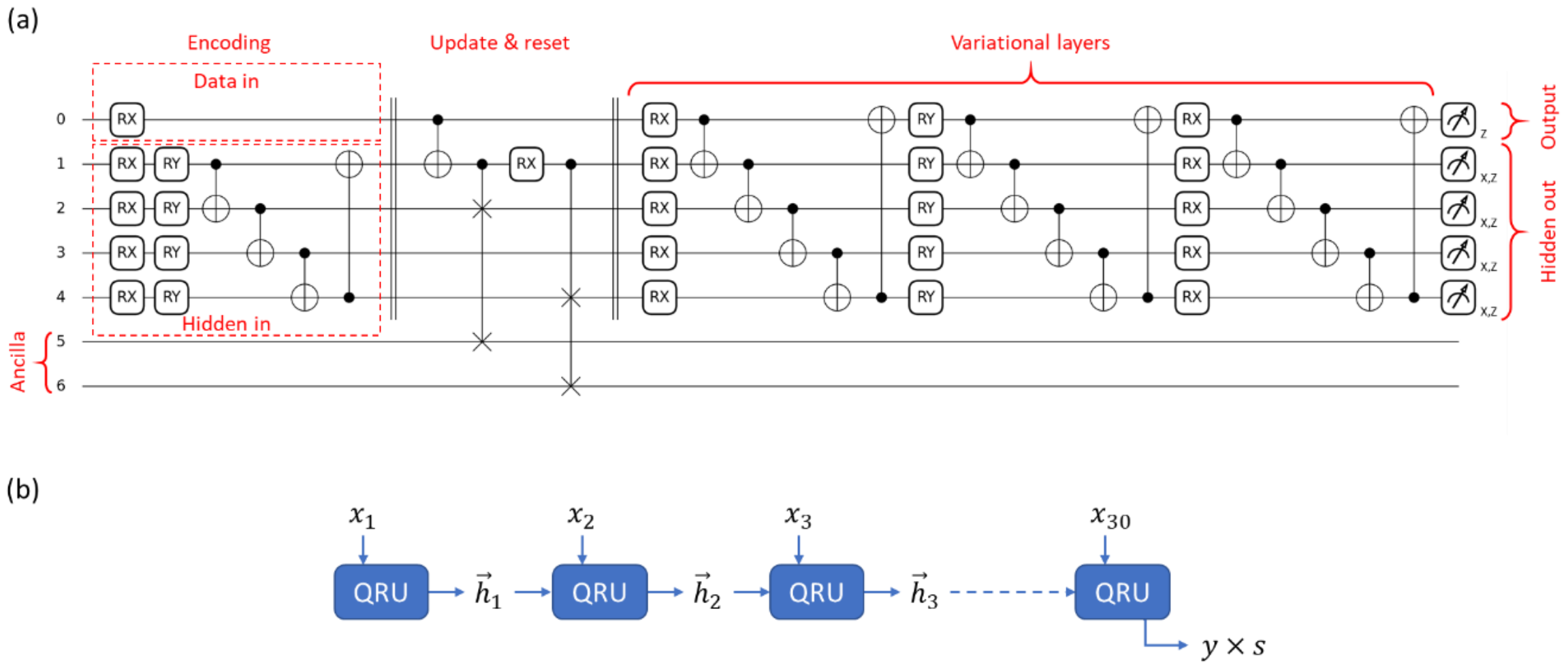

Figure S3: QRU circuit implementation for WDBC classification. (a) Circuit architecture with 1 data qubit and 4 hidden qubits. (b) Sequential processing of 30-dimensional feature vectors.

### S2.1 Circuit Architecture

The WDBC classification experiment uses a QRU configuration adapted for high-dimensional feature processing (Figure S3(a)). The circuit employs 1 data qubit for input encoding and 4 hidden qubits for temporal state representation, providing an 8-dimensional hidden state space.

Input and hidden state encoding follow the same angle encoding principles described in S1.1, with Rx and Ry rotation gates processing sequential feature inputs and hidden state propagation.

The architectural difference lies in the variational layers, which consist of 3 layers with alternating Rx-Ry-Rx rotation gates, each gate containing 1 trainable parameter. This alternating gate pattern provides sufficient expressivity for distinguishing between benign and malignant tumor features while maintaining parameter efficiency.

Output generation measures the Z-basis expectation value of the first qubit at the final processing step, with an additional scaling parameter *s* applied for binary classification optimization.

### S2.2 Sequential Processing Framework

The experiment processes the 30-dimensional WDBC feature vector as a sequential input ($x_1 - x_{30}$), treating each feature as a time step in a temporal sequence (Figure S3(b)). This approach allows the QRU model's recurrent architecture to handle high-dimensional data through its inherent sequential processing capability.

Each of the 30 features (extracted from digital images of breast tumor cell nuclei) is fed sequentially into the single data qubit, with the model maintaining temporal dependencies through its hidden state mechanism ($\vec{h}_i$). The raw output is scaled by a trainable parameter $s$ ($y \times s$), then processed through a sigmoid function for binary classification optimization. Training uses binary cross-entropy loss for malignant/benign tumor classification.

## S2.3 Training Convergence

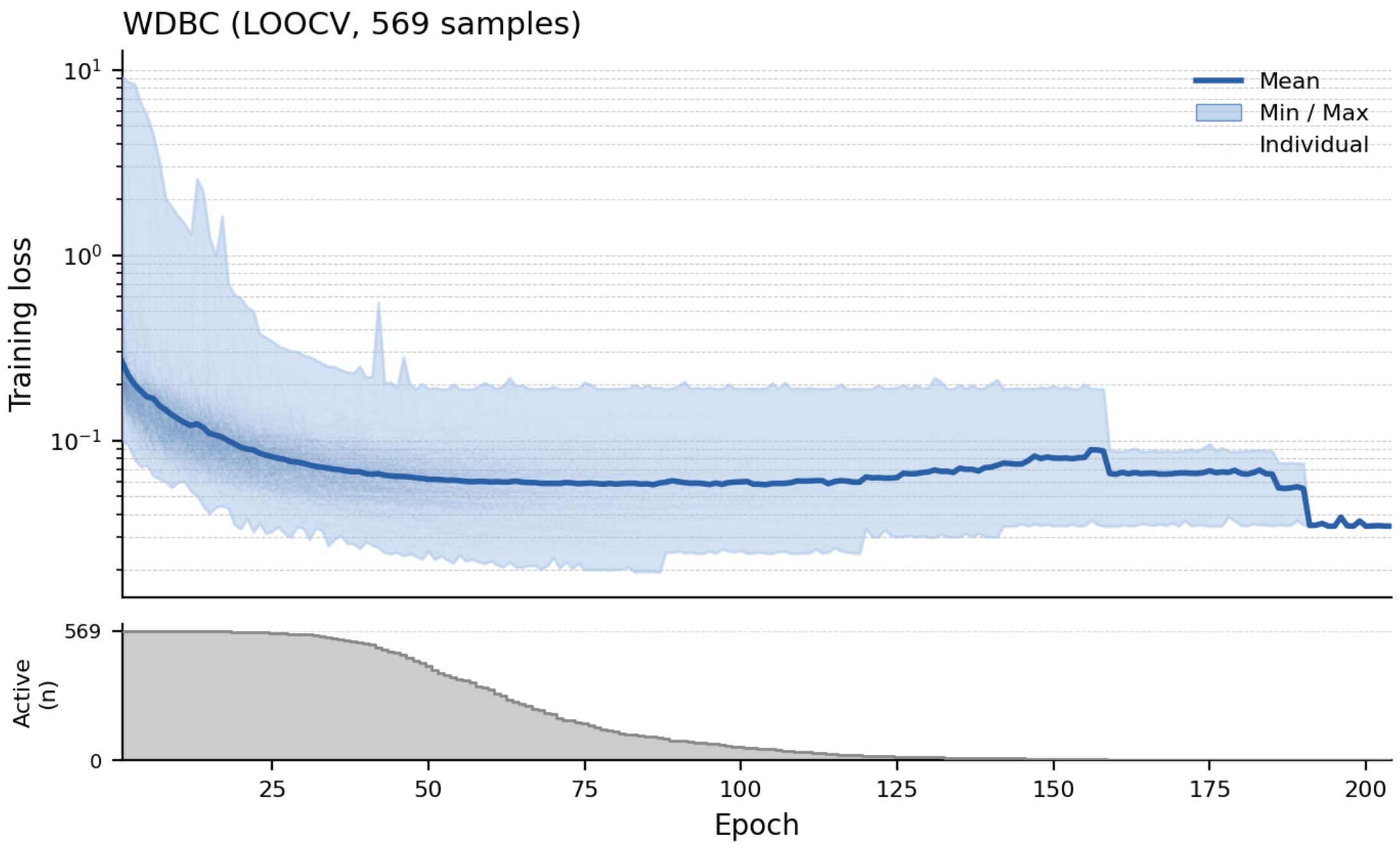

Figure S4: Training loss curves for the WDBC classification task across 569 LOOCV folds. Upper panel: mean training loss (solid line) with min/max range (shaded region) and individual runs (dashed lines), plotted on a logarithmic scale. Lower panel: number of active folds remaining at each epoch (early stopping applied individually per fold).

Training convergence was evaluated across all 569 Leave-One-Out Cross-Validation (LOOCV) folds. As shown in Figure S4, the mean training loss decreases rapidly in the first 25 epochs from approximately 2 to approximately 0.08, after which it stabilizes and continues to decrease gradually through the remaining epochs, converging to approximately 0.03 by the end of training. The wide min/max range reflects the natural variability across LOOCV folds arising from differences in the left-out sample and the resulting training set composition, rather than instability in the optimization process. The smooth, monotonic decrease in active fold count (lower panel) confirms that early stopping is applied consistently and that no folds exhibit divergent behavior. No evidence of barren plateaus or systematic trainability failure is observed.

## S2.4 Technical Specifications

- Total qubits: 7 (1 data + 4 hidden + 2 ancilla)

- Total trainable parameters: 35 (input encoding 2; hidden-state encoding 16; update-and-reset rotation 1; variational layers 15; scaling parameter s 1)
- Input sequence length: 30 features
- Output dimension: 1 (binary classification score)
- Hidden state dimension: 8 (4 qubits × 2 bases)
- Evaluation method: Leave-One-Out Cross-Validation (LOOCV)

## S3. MNIST Digit Recognition Implementation

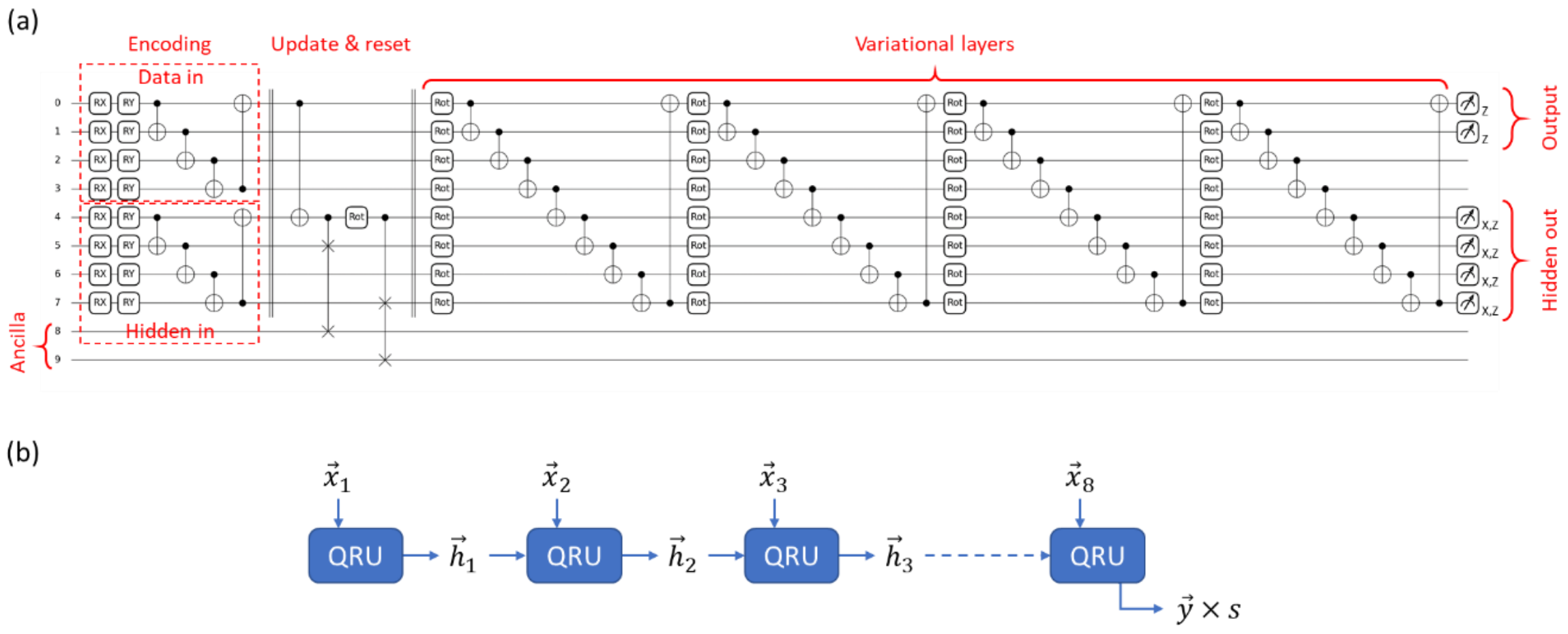

Figure S5: QRU circuit implementation for MNIST digit recognition. (a) Circuit architecture with 4 data qubits and 4 hidden qubits. (b) Sequential processing of 8×8 images over 8 time steps.

### S3.1 Circuit Architecture

The MNIST digit recognition experiment demonstrates QRU's capability in processing high-dimensional image data (Figure S5(a)). The circuit uses 4 data qubits for input encoding and 4 hidden qubits for temporal state representation, providing an 8-dimensional hidden state space.

Input and hidden state encoding follow the same angle encoding framework as described in S1.1, with the key difference being the increased data qubit count to handle higher-dimensional input vectors.

The variational layers consist of 4 layers using Rot gates (three-parameter single-qubit rotations implementing Rz($\omega$)Ry($\vartheta$)Rz($\phi$)), each containing 3 trainable parameters. This configuration provides enhanced expressivity suitable for the complexity of image recognition tasks.

Output generation measures the Z-basis expectation values of the first two qubits at the final processing step, creating a 2-dimensional output vector that is scaled by parameter *s* and processed through softmax for binary classification between digits "3" and "5".

### S3.2 Sequential Processing Framework

The experiment processes 8×8 pixel images through a sequential temporal framework (Figure S5(b)). Each 8×8 image is treated as a sequence of 8 time steps ($\vec{x}_1 - \vec{x}_8$), where each time step inputs one row of 8 pixel values. This row-by-row processing approach converts 2D spatial information into 1D temporal sequences, allowing the QRU model's recurrent architecture to handle image data through its natural sequential processing capability.

The 4 data qubits enable processing of 8 features per time step, with each qubit encoding 2 pixel values through Rx and Ry rotations. After 8 time steps, the complete image has been processed sequentially. The 2-dimensional raw output is scaled by a trainable parameter $s$ ($\vec{y} \times s$), then processed through a softmax function to generate probability distributions for the two digit classes ("3" and "5"). Training uses cross-entropy loss for classification optimization.

This sequential image processing demonstrates the QRU model's flexibility in handling non-temporal data by leveraging its recurrent structure to process spatial information through temporal decomposition.

### S3.3 Training Convergence

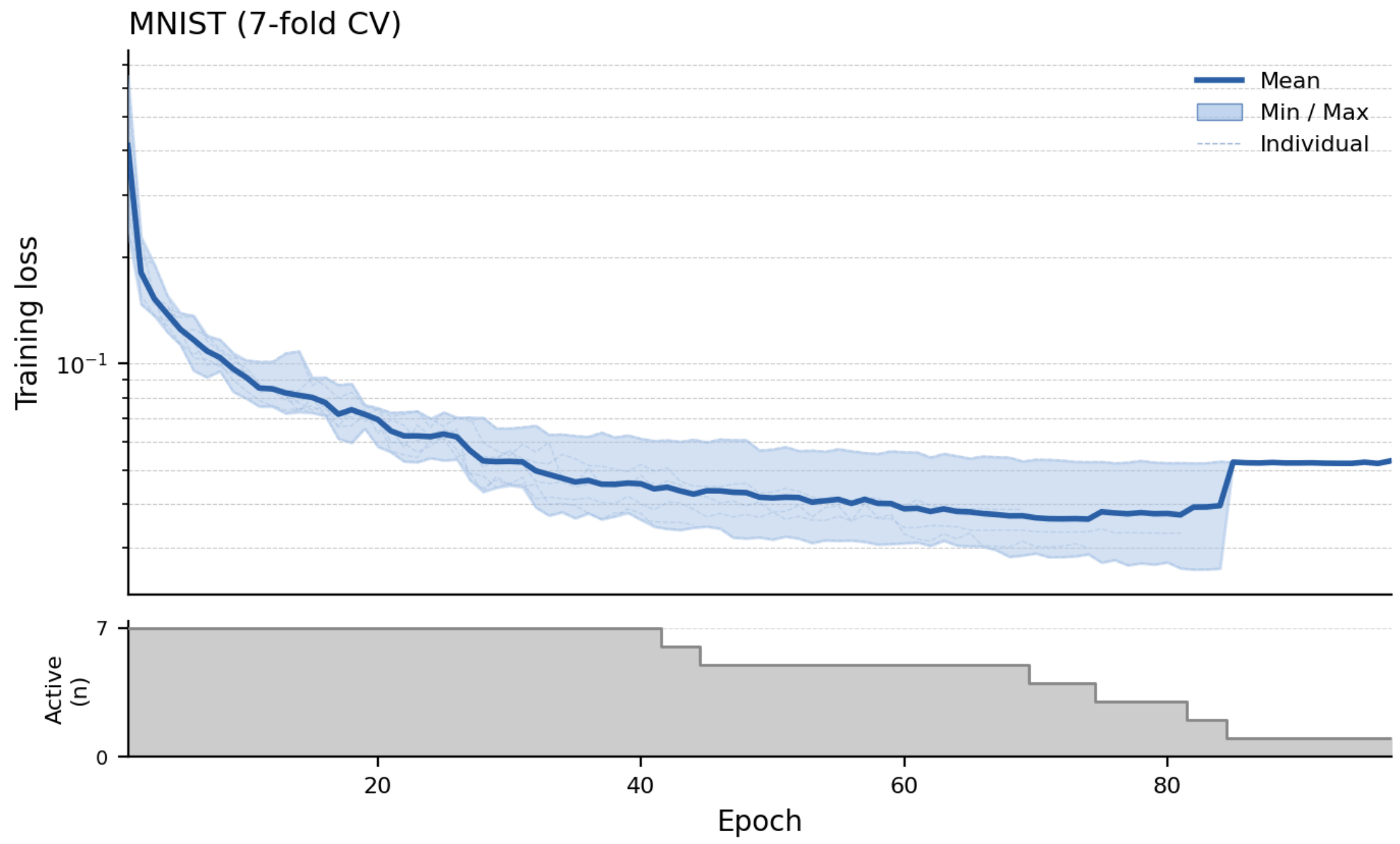

Figure S6: Training loss curves for the MNIST digit recognition task across 7 cross-validation folds. Upper panel: mean training loss (solid line) with min/max range (shaded region) and individual runs (dashed lines), plotted on a logarithmic scale. Lower panel: number of active folds remaining at each epoch (early stopping applied individually per fold).

Training convergence was evaluated across 7 stratified cross-validation folds. As shown in Figure S6, the mean training loss decreases rapidly in the first 20 epochs and converges to a stable plateau of approximately 0.04–0.05 by epoch 30, remaining stable thereafter. The narrow min/max range throughout training indicates high consistency across folds, reflecting the stability of QRU's optimization on this task. The stepwise decrease in the active fold count (lower panel) corresponds to the sequential early stopping of individual folds as training progresses. No evidence of barren plateaus or systematic trainability failure is observed across any of the 7 folds.

### S3.4 Technical Specifications

- Total qubits: 10 (4 data + 4 hidden + 2 ancilla)
- Total trainable parameters: 132 (input encoding 16; hidden-state encoding 16; update-and-reset rotation 3; variational layers 96; scaling parameter s 1)
- Input sequence length: 8 time steps
- Input dimension per time step: 8 pixel values
- Output dimension: 2 (binary classification vector)
- Hidden state dimension: 8 (4 qubits × 2 bases)
- Image size: 8×8 pixels
- Evaluation method: Stratified 7-fold cross-validation

## S4. Ablation Studies

### S4.1 Dual-Basis (Z+X) Measurement

The dual-basis measurement strategy was adopted to maximize the amount of quantum state information available for hidden state propagation within the constraint of a finite qubit count. At each time step, measuring each hidden qubit in both the Z and X bases provides a richer representation of the quantum state than single-basis measurement alone, at the cost of a modest increase in parameters for re-encoding the extended hidden state vector.

We compared two configurations on the damped oscillation prediction task over 10 independent runs:

  - Z-only: Hidden state derived from Z-basis measurements only; 4 hidden qubits providing a 4-dimensional hidden state; 21 trainable parameters.
  - Z+X: Hidden state derived from both Z- and X-basis measurements; 4 hidden qubits providing an 8-dimensional hidden state; 29 trainable parameters.

Both configurations represent an early design of QRU in which the entanglement topology differs slightly from the final architecture shown in Figure 2 of the main text; the measurement basis is the primary design variable under comparison. The circuit diagrams for both configurations are shown in Figure S7.

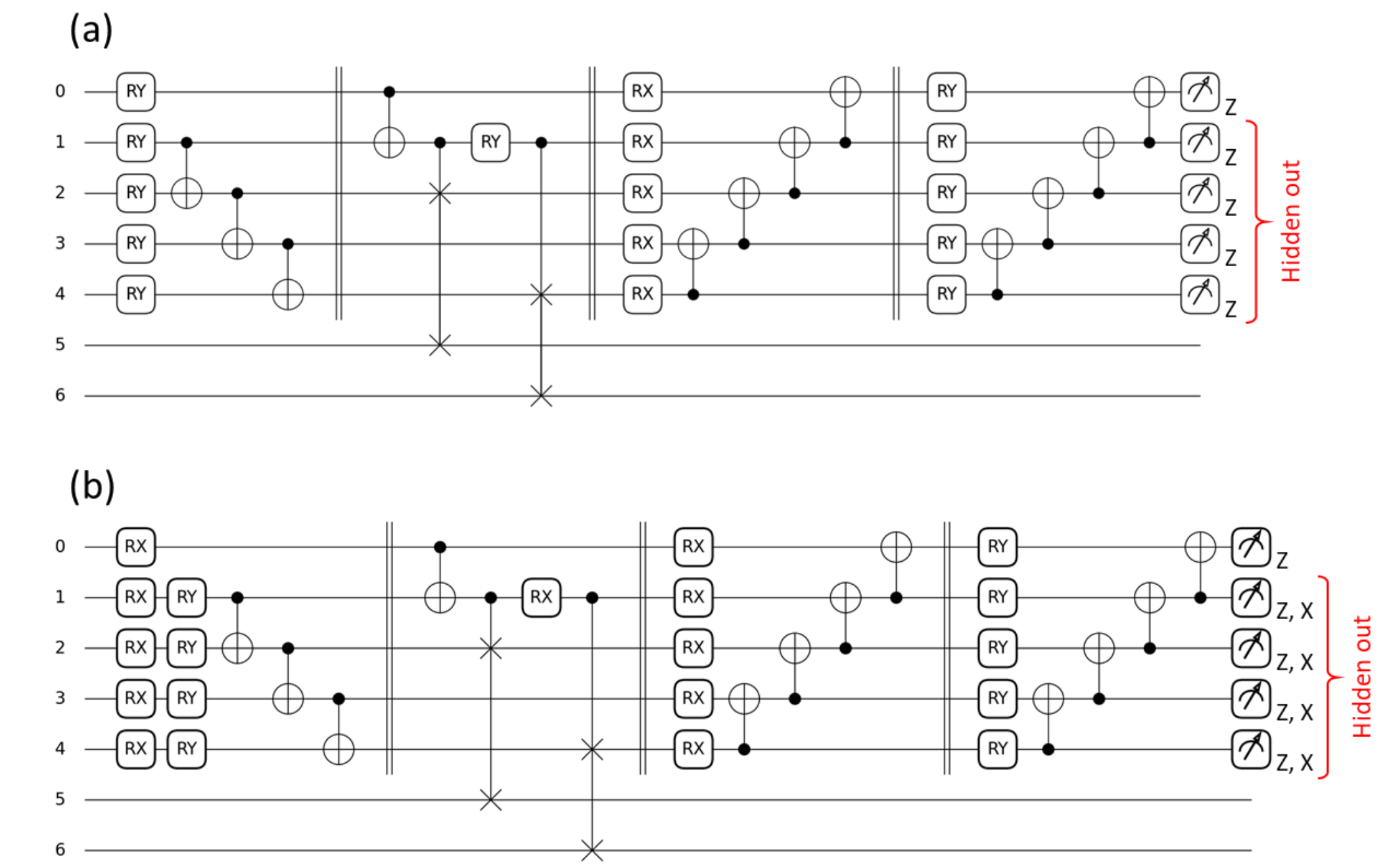

Figure S7: (a) Z-only measurement configuration: hidden state derived from Z-basis measurements only (21 trainable parameters, 4-dimensional hidden state). (b) Z+X dual-basis measurement configuration: hidden state derived from both Z- and X-basis measurements (29 trainable parameters, 8-dimensional hidden state). Both circuits represent an early design of QRU in which the entanglement topology differs slightly from the final architecture shown in Figure 2 of the main text.

Table S1: Test MSE comparison between Z-only and Z+X measurement configurations (10 independent runs).

| Configuration | Params | Hidden state dim | Test MSE Mean | Test MSE Std |
|---|---|---|---|---|
| Z-only | 21 | 4 | $1.699\times10^{-2}$ | $1.027\times10^{-2}$ |
| Z+X (dual-basis) | 29 | 8 | $2.260\times10^{-3}$ | $3.171\times10^{-3}$ |

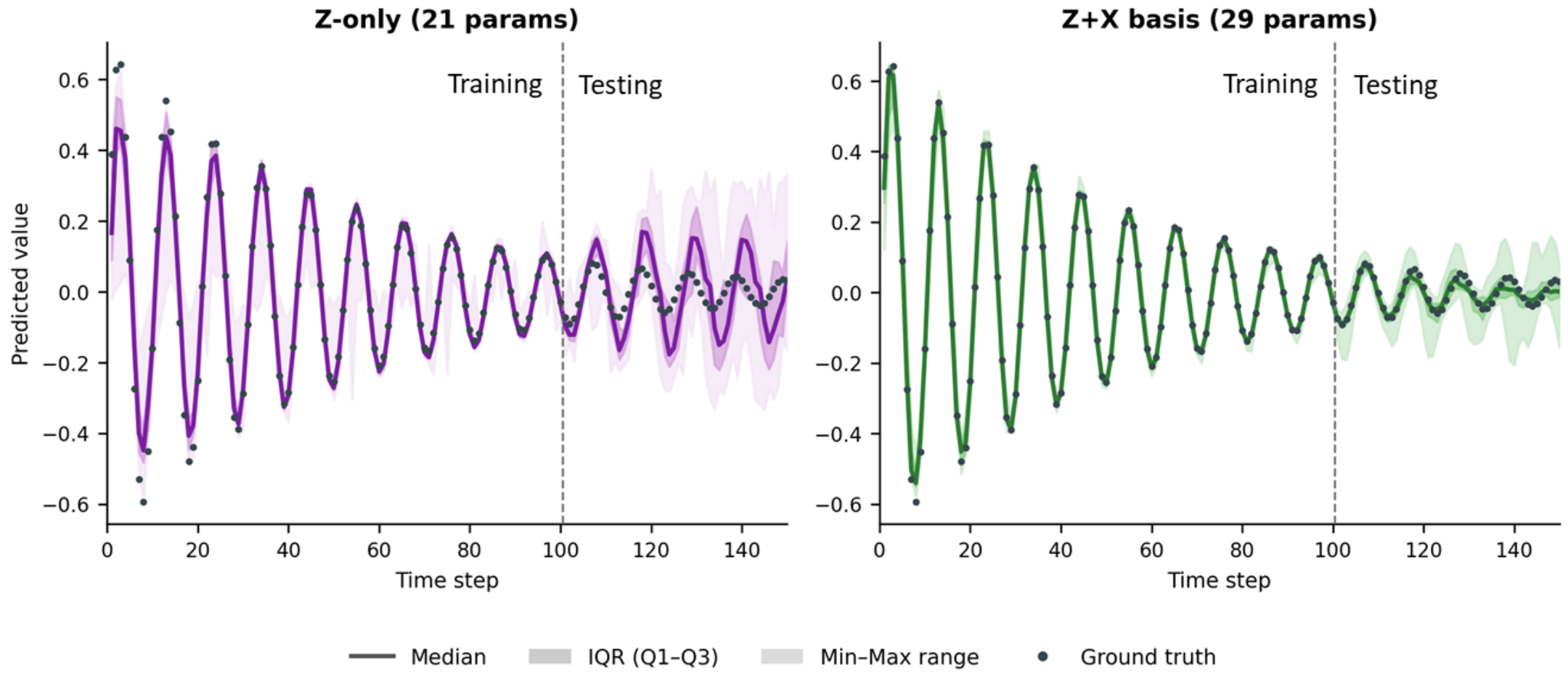

Figure S8: Prediction curves for the Z-only and Z+X measurement configurations on the damped oscillation task (10 independent runs each). Solid line: median run. Dark

shading: interquartile range (Q1–Q3). Light shading: full min–max range. Black dots: ground truth values. Dashed vertical line separates the training phase (steps 1–100) from the prediction phase (steps 101–150).

The dual-basis configuration yields an 86.7% reduction in test MSE and a substantially lower standard deviation, indicating improved accuracy and training stability (Table S1, Figure S8). This result confirms that the additional expressiveness from dual-basis measurements justifies the modest parameter increase, and validates the adoption of the Z+X strategy in the final QRU architecture.

The dual-basis design also reflects the original intent of QRU to propagate as much quantum state information as possible between time steps. As discussed in Section 2.2.6 of the main text, the measurement-and-re-encoding scheme is a NISQ-imposed compromise; the Z+X strategy represents a principled approach to maximizing information retention within this constraint.

## S4.2 C-SWAP Mechanism

To isolate the contribution of the C-SWAP-based information selection mechanism, we compared three architectural configurations under identical experimental conditions. All configurations use the dual-basis (Z+X) measurement scheme, 4 hidden qubits (providing an 8-dimensional hidden state), and 29 trainable parameters, and were evaluated on the damped oscillation prediction task (the same task as Section 3.2 of the main text). Ten independent training runs were performed for each configuration.

Configuration descriptions (Figure S9):

- Standard QRU: The full QRU architecture with C-SWAP operations for quantum-domain information selection (1 data qubit, 4 hidden qubits, 2 ancilla qubits; 7 qubits total).
- Without C-SWAP: The C-SWAP operations and ancilla qubits are removed. The hidden state is updated solely through the variational circuit without explicit information routing (1 data qubit, 4 hidden qubits; 5 qubits total).
- CNOT Replacement: The C-SWAP operations are replaced by CNOT gates, maintaining the same qubit structure as Without C-SWAP but with a simpler classical-equivalent entangling operation (1 data qubit, 4 hidden qubits; 5 qubits total).

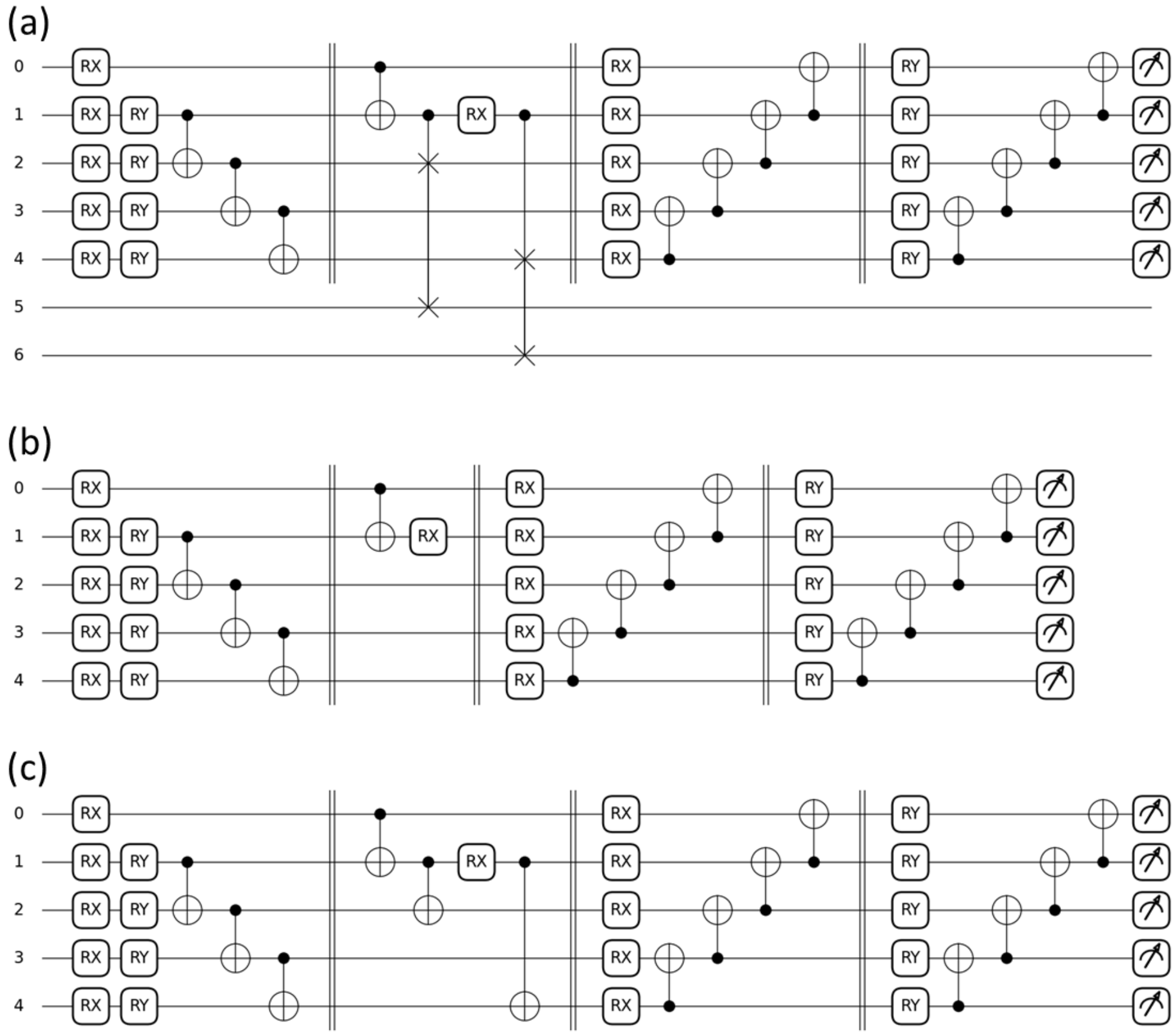

Figure S9: Circuit diagrams for the three C-SWAP ablation configurations. (a) Standard QRU: full architecture with C-SWAP operations (1 data qubit, 4 hidden qubits, 2 ancilla qubits; 7 qubits total; 29 trainable parameters). (b) Without C-SWAP: C-SWAP operations and ancilla qubits removed (5 qubits total). (c) CNOT Replacement: C-SWAP replaced by CNOT gates (5 qubits total). All configurations use dual-basis (Z+X) measurement and identical parameter budgets.

Table S2: Test MSE statistics for C-SWAP ablation configurations (steps 101–150, 10 independent runs).

| Configuration | Mean | Median | Std |
|---|---|---|---|
| Standard QRU (with C-SWAP) | $2.26\times10^{-3}$ | $7.4\times10^{-4}$ | $3.34\times10^{-3}$ |
| Without C-SWAP | $1.627\times10^{-2}$ | $1.474\times10^{-2}$ | $1.161\times10^{-2}$ |
| CNOT Replacement | $7.18\times10^{-3}$ | $4.12\times10^{-3}$ | $8.11\times10^{-3}$ |

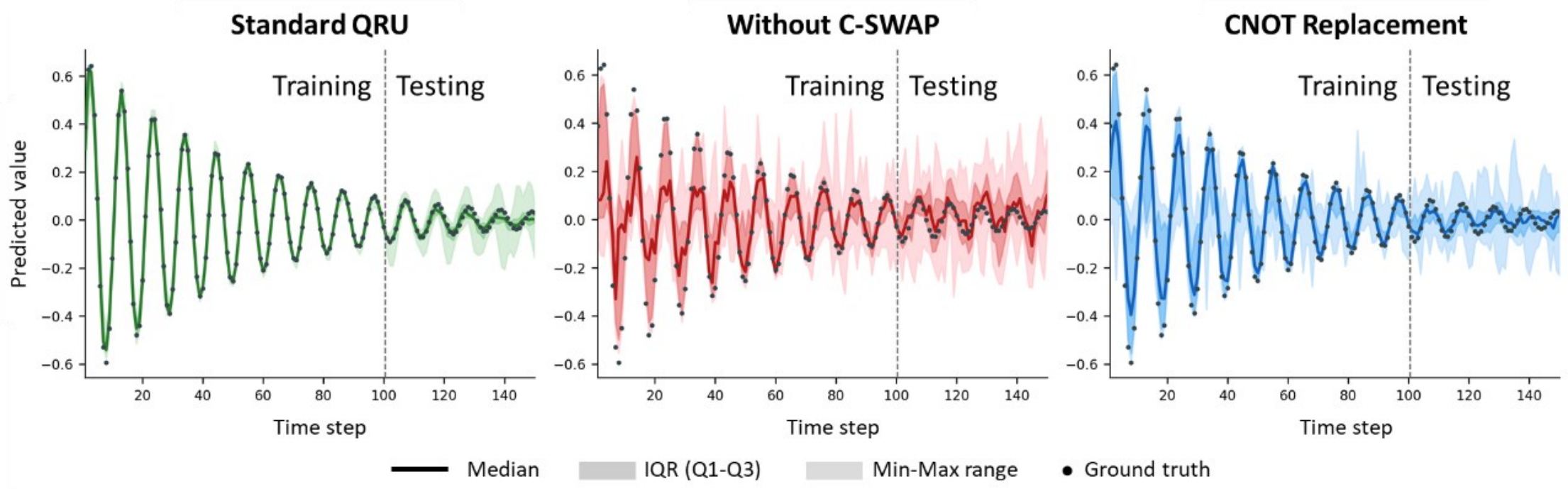

Figure S10: Prediction curves for the three C-SWAP ablation configurations on the damped oscillation task (10 independent runs each). Solid line: median run. Dark shading: interquartile range (Q1–Q3). Light shading: full min–max range. Black dots: ground truth values. Dashed vertical line separates the training phase (steps 1–100) from the prediction phase (steps 101–150).

The Standard QRU achieves a 20-fold lower median test MSE than the Without C-SWAP configuration and a 5.6-fold improvement over CNOT Replacement. The Without C-SWAP configuration also exhibits the highest variance across runs, indicating that removing the C-SWAP mechanism reduces not only predictive accuracy but also training stability. These results confirm that C-SWAP-based quantum-domain information routing is a critical and non-redundant architectural component (Table S2, Figure S10).

## S4.3 Input Feature Ordering Robustness (WDBC)

The 30 diagnostic features in the WDBC dataset do not carry a natural sequential ordering. To verify that QRU's classification performance is not sensitive to the choice of feature ordering, we conducted experiments with 5 different random feature permutations. Each permutation independently randomized both the feature ordering and the dataset split (340 train / 113 validation / 116 test, stratified by class label), with 10 independent training runs per permutation. The training protocol follows the same procedure as Section 3.3.1 of the main text, using a stratified random split in place of Leave-One-Out Cross-Validation. The results are summarized in Table S3.

Table S3: WDBC classification accuracy across 5 random feature permutations (10 runs each).

| Permutation | Mean Accuracy | Std |
|---|---|---|
| 1 | 95.34% | 0.69% |
| 2 | 95.69% | 1.85% |
| 3 | 92.93% | 1.72% |
| 4 | 94.31% | 2.50% |
| 5 | 94.91% | 1.98% |
| Overall | 94.64% | 0.97% (across permutation means) |

Since each permutation varies both the feature ordering and the dataset split simultaneously, the overall standard deviation of 0.97%—computed across the five per-permutation mean accuracies—reflects variability attributable to changes in feature ordering and dataset split rather than to within-permutation training randomness, providing a broad measure of robustness. The consistently high accuracy across all configurations confirms that QRU does not rely on any assumed sequential structure in the WDBC feature set, and that the C-SWAP-based information selection mechanism operates effectively regardless of input ordering.

## S5. Observable-Targeted Calibration Parameters

This section provides estimates of the Observable-Targeted Calibration (OTC) parameters used in the noise injection (NI) training strategy described in Section 3.5.1 of the main text. Both the noise model (NM) and fake backend (Fake) calibrations were performed using ibm_marrakesh as the target backend.

For each single-qubit observable, the signal attenuation coefficient $\alpha$ in Eq. (10) of the main text is estimated by ordinary least squares regression through the origin on the calibration pairs $\left(O_i^{\text{ideal}}, O_i^{\text{noisy}}\right)$, where *i* indexes the 12 calibration data points (4 target values, ±0.8 and ±0.4, × 3 repetitions):

$$\hat{\alpha} = \frac{\sum_i O_i^{\text{ideal}} \cdot O_i^{\text{noisy}}}{\sum_i \left(O_i^{\text{ideal}}\right)^2}.$$

The noise amplitude coefficient $K$ is estimated as:

$$\hat{K} = \frac{\text{std}\left(O_i^{\text{noisy}} - \hat{\alpha} O_i^{\text{ideal}}\right)}{\sqrt{\langle 1 - \hat{\alpha}^2 \left(O_i^{\text{ideal}}\right)^2 \rangle}},$$

where $\langle \cdot \rangle$ denotes the mean over all 12 calibration data points. This estimator follows from the moment condition $\mathbb{E}[\text{Var(residuals)}] = K^2 \mathbb{E}\left[1 - \alpha^2 \left(O^{\text{ideal}}\right)^2\right]$ implied by the noise model in Eq. (10) of the main text.

For the WDBC task, qubit 0 is the output qubit and qubits 1–4 are hidden qubits. For the MNIST task, qubits 0 and 1 are output qubits and qubits 4–7 are hidden qubits. Observables are listed in order of qubit index; within each qubit, the Z-basis observable precedes the X-basis observable. Parameters are calibrated independently for each observable and applied individually during NI training. The calibrated parameters are listed in Tables S4 and S5 for the WDBC and MNIST tasks, respectively.

Table S4: OTC calibration parameters for the WDBC task. NM = noise model (ibm_marrakesh); Fake = FakeMarrakesh.

| **Observable** | **NM $\alpha$** | **NM $K$** | **Fake $\alpha$** | **Fake $K$** |
|---|---|---|---|---|
| Z0 | 0.9230 | 0.0258 | 0.7362 | 0.0216 |
| Z1 | 0.9213 | 0.0222 | 0.7770 | 0.0269 |
| X1 | 0.9357 | 0.0337 | 0.8200 | 0.0364 |
| Z2 | 0.9270 | 0.0250 | 0.7179 | 0.0269 |
| X2 | 0.9277 | 0.0232 | 0.7792 | 0.0200 |
| Z3 | 0.8851 | 0.0221 | 0.7451 | 0.0227 |
| X3 | 0.9320 | 0.0246 | 0.7905 | 0.0390 |
| Z4 | 0.9067 | 0.0447 | 0.6979 | 0.0267 |
| X4 | 0.8872 | 0.0483 | 0.7360 | 0.0268 |

Table S5: OTC calibration parameters for the MNIST task. NM = noise model (ibm_marrakesh); Fake = FakeMarrakesh.

| Observable | NM $\alpha$ | NM $K$ | Fake $\alpha$ | Fake $K$ |
|---|---|---|---|---|
| Z0 | 0.8762 | 0.0270 | 0.5786 | 0.0513 |
| Z1 | 0.9059 | 0.0240 | 0.6852 | 0.0783 |
| Z4 | 0.8466 | 0.0294 | 0.5659 | 0.0410 |
| X4 | 0.8681 | 0.0201 | 0.5545 | 0.0727 |
| Z5 | 0.8707 | 0.0268 | 0.5231 | 0.0491 |
| X5 | 0.9294 | 0.0302 | 0.7709 | 0.0533 |
| Z6 | 0.8744 | 0.0227 | 0.5831 | 0.0326 |
| X6 | 0.9410 | 0.0456 | 0.6925 | 0.0323 |
| Z7 | 0.8575 | 0.0342 | 0.5365 | 0.0611 |
| X7 | 0.9230 | 0.0244 | 0.6953 | 0.0265 |

The NM calibration yields consistently higher $\alpha$ values than FakeMarrakesh for both tasks (WDBC: mean 0.916 vs. 0.756; MNIST: mean 0.889 vs. 0.619), reflecting a greater net reduction in observable magnitude due to gate noise on the fake backend because the fake backend model additionally considers ibm_marrakesh's hardware constraints such as qubit topology, coupling map, and basis gates. The difference is more pronounced for the MNIST task, consistent with the deeper transpiled circuit depth of the MNIST configuration.

## S6. Hardware-Aware Configuration Diagnostic Check

This section reports a diagnostic experiment designed to assess whether the mechanisms proposed in Section 3.5.3 contribute to the discrepancy observed between the noise model (NM), FakeMarrakesh, and readout-mitigated real-hardware results in Table 10.

The NM and FakeMarrakesh simulations used throughout Sections 3.5.1–3.5.3 employ Qiskit's default simulator settings. As discussed in Section 3.5.3, these default settings do not enforce identical qubit routing constraints between NM and FakeMarrakesh, and neither includes idle-time decoherence unless an explicit circuit scheduling pass is applied. To test whether these factors contribute substantially to the discrepancy observed in Table 10, we repeated the single-step OTC observable comparison (Section 3.5.3) under a hardware-aware configuration designed to more closely match real-device execution conditions.

Relative to the default settings used elsewhere in this work, three changes were made: the NM simulator was constructed with the full target object of FakeMarrakesh, so that the qubit topology, coupling map, native basis gates, and gate durations are explicitly enforced during transpilation, matching the FakeMarrakesh simulation; both the NM and FakeMarrakesh circuits were transpiled with an explicit scheduling pass, so that idle-time decoherence during periods when other qubits are being operated on is represented in the noise simulation; and the transpiler seed was fixed to ensure reproducibility of the qubit routing decisions. This configuration substantially increases simulation cost relative to the default settings, since scheduling requires explicit gate-duration information and produces deeper, more constrained circuits. For this reason, it is used here only as a single-step diagnostic check and was not applied to the noise-injection training experiments in Sections 3.5.1–3.5.2, which require repeated simulation across many training iterations.

Table S6 compares the mean deviation from the statevector reference (target expectation value ≈ +0.8 group, following the convention used in Table 10) under the default and hardware-aware configurations.

Table S6: OTC observable comparison under default and hardware-aware simulation settings.

| Configuration | Noise Model | FakeMarrakesh |
|---|---|---|
| Default | -13.0% | -38.1% |
| Hardware-aware | -59.7% | -59.8% |

(For reference, the readout-mitigated real-hardware deviation reported in Table 10 is −78.2%.)

Under the hardware-aware configuration, the NM and FakeMarrakesh deviations converge closely, consistent with qubit topology and coupling-map asymmetry contributing substantially to disagreement between the two simulations under default settings. Both deviations also move substantially closer to the readout-mitigated real-hardware result, consistent with idle-time decoherence contributing meaningfully to the gap between simulated and real-hardware noise. The remaining difference from the real-hardware result may arise from the fact that simplified backend-derived noise models do not fully capture more complex hardware noise mechanisms such as crosstalk, correlated or nonlocal errors, non-Markovian noise dynamics, etc., as discussed in Section 3.5.3.

## S7. QLSTM/QGRU Comparison (MNIST)

To provide a comparison between QRU and representative QLSTM/QGRU implementations from the literature, complementing the paradigm distinction discussed in Related Work (Section 2.1.4), we implemented and trained representative QLSTM/QGRU designs from the literature — Ceschini et al.'s (2024) QGRU and QLSTM reformulation [26], and Chen et al.'s (2022) original QLSTM [25] — on the MNIST "3" vs "5" classification task, under the reduced-cost protocol of Table 9, Section 3.5 (identical first-fold split of the 7-fold cross-validation, raw output without a trainable scaling parameter, MSE loss, single run), rather than the full protocol behind the 98.05% reported in Section 3.4.2. All models, including QRU, were trained in single (float32) precision; QRU's accuracy in Table S7 (97.19%) therefore differs slightly from its double-precision value in Table 9 (97.09%).

Table S7: Parameter-matched comparison between QRU and representative QLSTM/QGRU implementations from the literature, on the MNIST "3" vs "5" task, under the reduced-cost protocol of Table 9, in single precision.

| Model | Circuit configuration | Trainable parameters | Test accuracy | Training epochs |
|---|---|---|---|---|
| QRU | 1 circuit, 10 qubits | 131 | 97.19% | 909 |
| QGRU (Ceschini et al., 2024) | 3 VQCs × (4 qubits, 3 layers) | 144 | 95.27% | 460 |
| QGRU (Ceschini et al., 2024) | 3 VQCs × (4 qubits, 4 layers) | 156 | 92.46% | 1,240 |
| QLSTM (Ceschini et al., 2024) | 4 VQCs × (4 qubits, 3 layers) | 156 | 95.99% | 900 |
| QLSTM (Ceschini et al., 2024) | 4 VQCs × (4 qubits, 4 layers) | 172 | 95.21% | 1,140 |
| QLSTM (Chen et al., 2022) | 4 VQCs × (10 qubits, 1 layer) + 2 VQCs × (2 qubits, 2 layers) | 144 | 53.07%* | 450 |
| QLSTM (Chen et al., 2022) | 4 VQCs × (10 qubits, 5 layers) + 2 VQCs × (2 qubits, 2 layers) | 624 | 68.99% | 350 |

*For Chen et al.'s design at one LSTM-gate VQC layer, training loss remained flat across the entire run regardless of learning-rate schedule, and the reported accuracy exactly matches the test set's majority-class ratio; the model did not learn.

Note: For QGRU, the three VQCs correspond to the reset, update, and candidate gates. For QLSTM (both Ceschini-style and Chen et al.), the four LSTM-gate VQCs correspond to the forget, input, candidate, and output gates. Chen et al.'s design additionally includes two smaller VQCs that map the combined cell state to the hidden state and output.

QLSTM and QGRU-style quantum recurrent architectures share a common paradigm: each replaces the classical weight matrices used to compute an LSTM/GRU-gate's pre-activation with a variational quantum circuit (VQC), while retaining the classical gating nonlinearities and elementwise state updates that combine the gates. Chen et al.'s design and Ceschini et al.'s reformulation are two representative instances of this paradigm, differing in how the previous hidden state and current input, concatenated as $[h_{t-1}, x_t]$ (following Ceschini et al.'s notation), are prepared for the VQC, and in how the VQC outputs are mapped back to the hidden state and final output. Chen et al. encode the concatenated state directly, one feature per qubit, with no classical compression, and use two additional VQCs to map the combined cell state to the hidden state and output, respectively. Ceschini et al. instead pass the concatenated state through a shared classical input-compression layer (denoted $FC_{in}$ in their architecture) before each LSTM/GRU-gate's VQC, and use a corresponding shared classical output layer ($FC_{out}$) to map the LSTM/GRU-gate VQCs' outputs back to the hidden state and output — eliminating the need for the two additional VQCs that Chen et al.'s design requires.

Because every LSTM/GRU-gate under this paradigm requires an independent parameterized VQC with sufficient circuit expressivity to train, designs following this paradigm incur this expressivity cost once per LSTM/GRU-gate rather than once overall. Ceschini et al.'s FC-based compression keeps this multiplicative cost cheap by decoupling per-LSTM/GRU-gate qubit count from raw input dimensionality; each of Chen et al.'s four LSTM-gate VQCs (forget, input, candidate, and output), which does not use this compression, requires as many qubits as the sum of the hidden and input dimensions — 10 qubits even at a hidden dimension of 2, the smallest tested for this task — compounding the multiplicity with a qubit-count burden that grows with task size. QRU avoids this multiplicative cost altogether: because gating is implemented natively via C-SWAP routing rather than by combining separately computed VQC outputs through classical nonlinearities, it requires only a single VQC circuit per timestep, not one per gate.

This structural cost is reflected directly in the parameter-matched comparison in Table S7. We tested Chen et al.'s design at one LSTM-gate VQC layer — chosen to approximate QRU's parameter budget (144 vs. 131, 1.1×) rather than to identify an optimized configuration for this design. At this layer count, Chen et al.'s design failed to train, converging to a constant majority-class prediction. Increasing LSTM-gate VQC depth to five layers (624 parameters) restored measurable training progress (68.99%), confirming that the single-layer failure reflects insufficient circuit expressivity for a 10-qubit circuit rather than a fundamentally untrainable architecture; this depth increase was intended only to demonstrate this trend, not to identify Chen et al.'s best achievable accuracy at any parameter budget.

Ceschini et al.'s QGRU and QLSTM reformulations reach test accuracies of 92.46%–95.99% with a parameter budget 1.1–1.3× QRU's. Neither this nor the Chen et al. comparison above is intended as a head-to-head performance claim against either literature design, particularly as each configuration was trained in a single run.

The structural cost of the QLSTM/QGRU paradigm discussed above arises from the classical gating logic that both designs retain (Section 2.1.4): each LSTM/GRU-gate is computed by its own separate VQC, and combining these separately-computed gate values — for example, via $c_t = f_t * c_{t-1} + i_t * \tilde{C}_t$ in Chen et al.'s QLSTM — requires each VQC's output as a definite classical number, and therefore measurement between every VQC and the next. QRU's single circuit has no such requirement, since it never combines separately measured outputs in the first place.

This is also why QRU's natural comparison targets throughout this work are classical recurrent models (Section 2.1.4), not QLSTM or QGRU: aiming from the outset toward a fully quantum recurrent form, QRU's long-term goal is to explore whether such an architecture can match or exceed classical recurrent models as hardware matures.

## S8. Error Mitigation on ibm_phoenix

PEC cancels the learned noise by sampling randomly modified versions of the circuit, at a cost in samples (the sampling overhead) that grows exponentially with the total learned noise of the circuit. On ibm_marrakesh, this circuit required 174 CZ gates (summed reported CZ error 0.42), and Qiskit Runtime reported a sampling overhead of at least $2.6 \times 10^5$; at Qiskit Runtime's default precision, this would require about $10^9$ shots, beyond the available QPU time. Thus, at the time of submitting the manuscript, we did not perform PEC on quantum hardware. During the manuscript review process, a reviewer requested an open and reproducible mitigation baseline in addition to the commercial Qiskit Function QESEM. So, we decided to employ PEC, which should be measured on another QPU with much less circuit sampling overhead, which grows exponentially with the circuit's total noise. The ibm_phoenix is a Nighthawk r2 QPU with broad cloud availability in September 2026 and is designed to improve circuit throughput by 25 times, marking it as IBM's fastest QPU to date. Its square-lattice connectivity reduces the circuit to 84 CZ gates (summed reported CZ error 0.12) and the sampling overhead to 10.8–11.8. PEC was therefore measured on ibm_phoenix, which became available after the other hardware measurements in Section 3.5.

A PEC run executes randomly generated circuit instances (randomizations) with a fixed number of shots per instance; both numbers are Qiskit Runtime options. In all runs, the number of randomizations executed was equal to the randomizations value we set multiplied by the sampling overhead learned in that run, rounded up (Table S8). The same transpiled circuit was measured in 15 independent runs, each with its own noise learning, with the number of randomizations set to 400, 800, or 1,600 and 128, 256, or 512 shots per randomization, using twirled readout error extinction, noise learning on all distinct layers, and no limit on the sampling overhead. Each run used $0.56–2.31 \times 10^6$ shots and 77–121 s of QPU time. The mean deviation from the statevector reference is −6.0%, with all runs below the reference (range −3.2% to −9.7%).

**Table S8: PEC runs on ibm_phoenix.** Values are listed per run. Overheads are those reported by Qiskit Runtime; Std is the reported statistical uncertainty, averaged over the five observables and the three runs.

| Randomizations | | | Shots per randomization | Mean deviation (%) | Std |
|---|---|---|---|---|---|
| Set value | Overhead | Executed (Set value × Overhead) | | | |
| 400 | 11.76, 11.00, 11.03 | 4703, 4402, 4413 | 128 | −3.5, −5.9, −4.3 | 0.020 |
| 400 | 10.81, 11.19, 11.01 | 4324, 4477, 4403 | 256 | −7.5, −6.3, −4.7 | 0.020 |
| 400 | 11.28, 10.88, 11.01 | 4514, 4354, 4403 | 512 | −3.2, −7.3, −7.0 | 0.020 |
| 800 | 11.14, 10.90, 11.05 | 8909, 8724, 8842 | 128 | −6.3, −9.7, −5.8 | 0.014 |
| 1,600 | 11.19, 11.05, 10.87 | 17898, 17681, 17387 | 128 | −5.3, −5.5, −8.3 | 0.010 |

To separate the effect of the device from that of the mitigation method, readout mitigation, gate twirling, and ZNE were also applied on ibm_phoenix to the transpiled circuit used for the PEC runs, with the settings of Table 10 and three runs each (Table S9). Readout mitigation and gate twirling deviate from the statevector reference by −43.8% and −38.0%, compared with −78.2% and −83.4% on ibm_marrakesh, reflecting the lower total noise of this circuit on ibm_phoenix. ZNE overshoots the reference, and in each ZNE setting, 7 of the 15 extrapolated expectation values (three runs × five observables) exceed the physical bound of 1. This reflects the nature of ZNE in artificial-noise amplification followed by curve fitting, whose validity depends on the choice of extrapolation ansatz (e.g., linear, exponential); ZNE is inherently biased. Therefore, on the same device (ibm_phoenix), our experiment results indicate that PEC, an unbiased estimator, remains much closer to the reference than the other methods.

Table S9: Readout mitigation, gate twirling, and ZNE on ibm_phoenix, applied to the same transpiled circuit as the PEC runs with the settings of Table 10 (three runs each). The mean deviation is computed as in Table 10.

| Method | Mean deviation from SV (%) | |
|---|---|---|
| | ibm_phoenix | ibm_marrakesh, Table 10 |
| Readout mitigation only | −43.8 | −78.2 |
| Twirling (300 inst.) | −38.0 | −83.4 |
| ZNE (32 twirling inst.) | +24.9 | −70.9 |
| ZNE (300 twirling inst.) | +29.8 | −65.2 |
| PEC | −6.0 | — |